\documentclass[reprint,pra,showpacs,amsmath,amssymb,superscriptaddress,showkeys]{revtex4-1}

\usepackage{bm}
\usepackage{float}
\usepackage{mathtools}
\usepackage{multirow}
\usepackage{bbold}
\usepackage{bbm}
\usepackage{braket}
\usepackage{bbold}
\usepackage{amsthm}
\usepackage{mathrsfs}
\usepackage[usenames]{xcolor}
\usepackage{gensymb}
\usepackage{appendix}
\usepackage{graphicx}
\usepackage[caption=false]{subfig}
\usepackage[normalem]{ulem}
\usepackage{color}

\usepackage{soul}
\usepackage{hyperref}

\DeclareMathOperator{\Tr}{Tr}

\def\L{\mathcal L}

\begin{document}
\def\mean#1{\left< #1 \right>}

\title{Driven Bose-Hubbard dimer under nonlocal dissipation: A bistable time crystal}

\author{C. Lled\'{o}}
\email[electronic address: ]{c.lledo.17@ucl.ac.uk}
\affiliation{Department of Physics and Astronomy, University College London,
Gower Street, London, WC1E 6BT, United Kingdom}
\author{Th. K. Mavrogordatos}
\affiliation{Department of Physics, Stockholm University, SE-106 91, Stockholm, Sweden}
\author{M. H. Szyma\'{n}ska}
\affiliation{Department of Physics and Astronomy, University College London,
Gower Street, London, WC1E 6BT, United Kingdom}

\date{\today}

\begin{abstract}
We investigate the critical behavior of the open coherently-driven Bose-Hubbard dimer under nonlocal dissipation. A conserved quantity arises from the nonlocal nature of the dissipation, rendering the dimer multistable. In the weak-coupling semiclassical limit, the displayed criticality takes the form of amplitude bistability and breaking of spatial and temporal symmetries. A period-bistable time crystal is formed, consisting of Josephson-like oscillations. Mean-field dynamics and quantum trajectories complement the spectral analysis of the Liouvillian in the approach to the semiclassical limit. 
\end{abstract}


\maketitle

\section{Introduction}
Phase transitions substantiated by a mean-field bifurcation have been intimately tied to the evolution of quantum optics. Characteristic cases include the Dicke phase transition \cite{Dicke}, co-operative resonance fluorescence \cite{NardPRA}, and the laser \cite{deGPRA}. The study of quantum phase transitions in zero dimensions \cite{PlenioQPT} far from thermal equilibrium reappraises the definition of the thermodynamic limit alongside the distinct role of quantum fluctuations, taking the form of an ostensible deviation from the mean-field dynamics \cite{SavageSA, Carmichael2015}. Furthermore, the work of Ref. \cite{Kessler2012} has already drawn the analogy between quantum phase transitions in closed systems and dissipative phase transitions (DPTs) in open quantum systems, through analyzing the spectrum of the Liouvillian, while the properties of the steady state have been assessed with respect to the particular critical eigenvalue defining the Liouvillian spectral gap \cite{CiutiSpectralTheo}. To date, DPTs attract a significant and ongoing interest among the communities of cold atoms, circuit QED, and semiconductors (see, e.g., Refs. \cite{Dimer2007, Baumann2010, Baumann2011, Brennecke2013,Fitzpatrick2017,Rodriguez2017, Fink2017, Fink2018}). 

In that background, the interplay between coherence and dissipation renders the Bose-Hubbard dimer (BHD) a suitable candidate for assessing the role of conserved quantities and the associated symmetries in dissipative quantum phase transitions. A transition from intermittent entanglement to a persistently entangled state was demonstrated a few years ago for a dissipative BHD \cite{EntangBHD}. The standard phenomenological approach to dissipation amounts typically to the definition of a Liouville superoperator governing the system response, assessed against its coherent evolution. Statistical independence of the reservoirs the system is coupled to gives rise to the configuration of {\it local dissipation}, while coupling to one common bosonic reservoir generates {\it nonlocal dissipation}.

Critical behavior in the presence of a coherent drive, leading to symmetry breaking, has been very recently taken up for local dissipation when characterizing the quantum correlation properties of the BHD \cite{CaoPRA, Casteels-Ciuti_PRA2017}. In many cases, local dissipation is well justified and often greatly simplifies the analysis of the transient and the steady state. In reality, however, different collective modes have different dissipation rates, giving rise to some amount of nonlocal dissipation.

In this work, we demonstrate the importance of the nonlocality of the dissipation in the open coherently-driven BHD. A continuous swapping symmetry together with a conserved quantity arises, leading to the occurrence of quantum multistability \cite{VictorAlbert}. Alongside the familiar first-order DPT \cite{Casteels-Fazio-Ciuti_PRA2017} associated with semiclassical amplitude bistability, with origins in the quantum Duffing oscillator \cite{Drummond1980}, the BHD breaks the spatial swapping symmetry as well as the time invariance of the steady state. By studying the closure of the Liouvillian gap as we approach a weak-coupling semiclassical limit, we show that a dissipative time-crystal \cite{SachaRPP, Keeling2018, Poletti2018, Pal2018, Rey2018, Ueda2018, Lesanovsky2019} with Josephson-like oscillations is formed. We find that two different time-crystalline periods coexist in the parameter regime of semiclassical bistability. We employ and compare the semiclassical Gross-Pitaevskii equations, quantum trajectories, and numerical solutions of the Master Equation in a truncated Hilbert space in order to fully appreciate the novel phases.
\begin{figure*}
\includegraphics[width=1\linewidth]{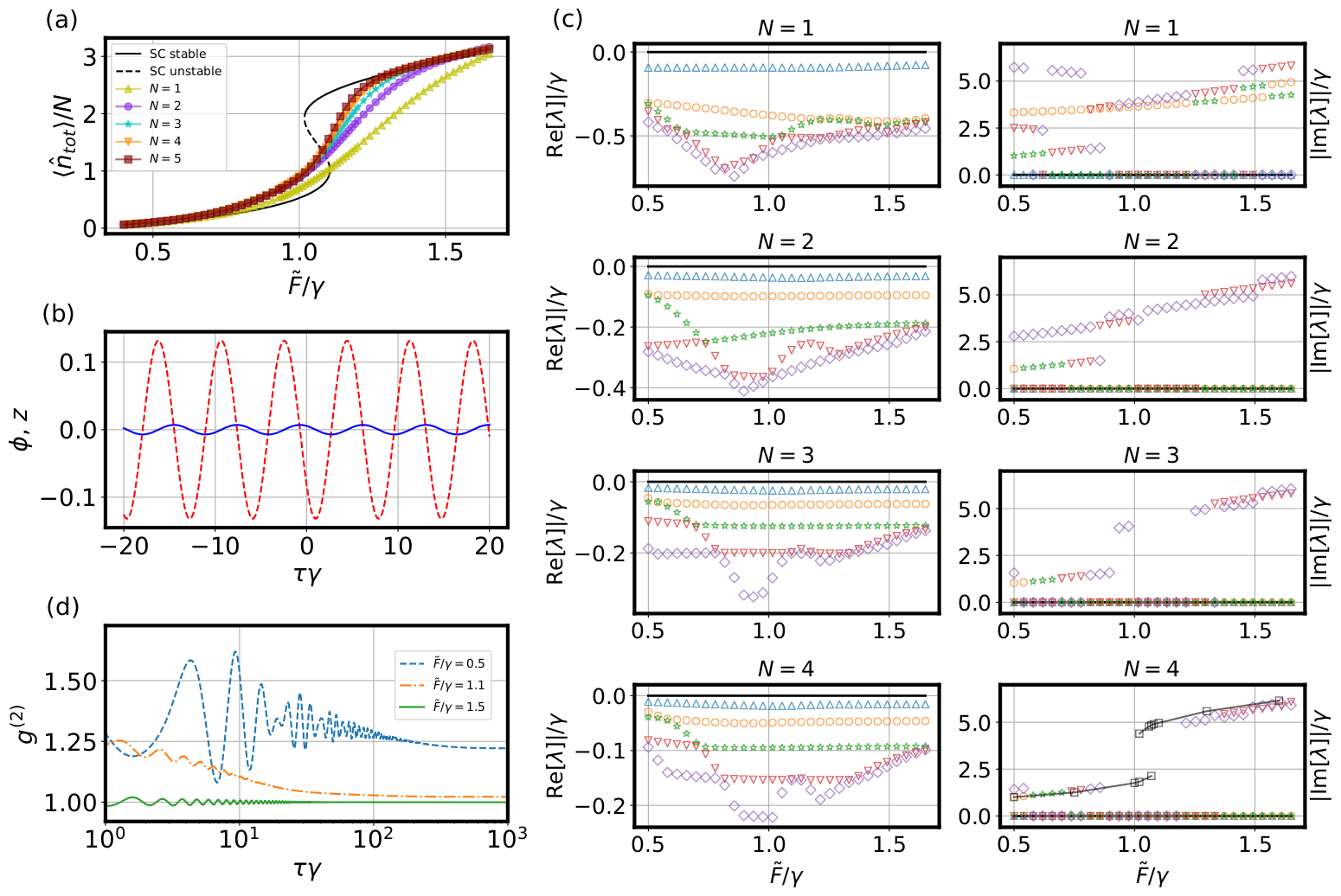} 
\caption{Limit cycles and the Liouvillian spectrum. (a), (c) Steady-state total number of bosons $\langle \hat n_\text{tot}\rangle/N$ and the first five Liouvillian gaps, respectively, as a function of $\tilde F/\gamma$ for different values of $N$. In (a) the black solid and dashed lines, denoted by {\it SC}, depict semiclassical predictions. In (c) the semiclassical limit-cycle frequencies are juxtaposed (in black squares) to the plot depicting the imaginary part of the gaps for $N=4$. (b) Relative phase $\phi = \text{Arg}[\alpha_1] - \text{Arg}[\alpha_2]$ (dashed red) and population difference $z= (|\alpha_1|^2-|\alpha_2|^2)/N$ (blue) versus time in the semiclassical Josephson limit cycle for $\tilde F/\gamma = 0.4$. The $40\gamma^{-1}$ time window is centered around $t_0=3.5\times 10^5 \gamma^{-1}$. (d) Quantum second-order correlation function $g^{(2)}(\tau)$ (symmetric with respect to the two modes $i=1,2$) for three different pump amplitudes in the quantum regime and $N=5$. The parameters are: $\Delta/\gamma = 0.7$, $J/\gamma=1.5$, and $\tilde U/\gamma = 1$. The Fock space per site is truncated at 10, 14, 18, 21, and 24 bosons for $N=1, 2, 3, 4$, and $5$, respectively.}
\label{fig: figure1}
\end{figure*}

\section{The model}
We consider a BHD whose dynamical evolution is governed by the Lindblad equation (with $\hbar=1$)
\begin{equation} \label{equ: Lindblad equation}
 \partial_t \hat \rho = -i[\hat{\mathcal H}, \hat \rho] + \gamma \mathcal D[\hat a_1 + \hat a_2](\hat \rho) \equiv \mathcal L(\hat \rho).
\end{equation}
Here, $\mathcal D[\hat L](\bullet) = \hat L \bullet \hat L^\dag - (1/2)\{\hat L^\dag \hat L, \bullet \}$ is the dissipator, while the system Hamiltonian under coherent drive, in a frame rotating with the pump frequency $\omega_p$, reads
\begin{equation} \label{equ: BH Hamiltonian rotating frame}
\begin{split}
 \hat{\mathcal H} = \sum\limits_{i=1,2}-\Delta & \hat a^\dag_i \hat a_i + U \hat a^\dag_i \hat a^\dag_i \hat a_i \hat a_i  + F(\hat a^\dag_i + \hat a_i) \\
&-J(\hat a^\dag_1 \hat a_2 + \hat a_{1} \hat a_{2}^\dag),
\end{split}
\end{equation}
where $\hat a_i$ is the $i$-mode bosonic annihilation operator, $\Delta=\omega_p - \omega_c$ is the detuning between the pump and the resonant mode frequencies, $\omega_p$ and $\omega_c$ respectively, $U$ is the interaction strength, $F$ is the driving amplitude, and $J$ is the intermode coupling. For nonlocal dissipation, $\hat L=\hat{a}_1 + \hat{a}_2$. The driving term originates from the interaction of the two bosonic modes $\hat a_{1,2}$ with a coherent source treated as a (semi)classical field. Drive and decay only affect the bonding mode $(\hat a_1 + \hat a_2)/\sqrt{2}$, which couples to the antibonding mode $(\hat a_1 - \hat a_2)/\sqrt{2}$ solely via the nonlinearity. We rescale the pump amplitude and the interaction with the help of a generic parameter $N$ (the analog of the laser saturation photon number, for instance), as $F = \sqrt{N} \tilde F$ and $U=\tilde U/N$, such that $\sqrt{U}F$ is constant for any $N$. This allows us to consider a well defined weak-coupling thermodynamic limit, $N\to \infty$, where the boson number diverges ($\langle\hat a_i^\dag \hat a_i\rangle \propto N$). In this limit, quantum fluctuations are negligible, and a semiclassical analysis is adequate \cite{Casteels-Ciuti_PRA2017}. Our system exhibits {\it three} important properties delineated below.

{\it The first one} is the presence of a continuous symmetry $\L(\hat U(\phi) \bullet \hat U(\phi)^\dag) = \hat U(\phi) \L(\bullet)\hat U(\phi)^\dag$ generated by the unitary operator $\hat U(\phi) = e^{i\phi \hat Z_2}$ (for real $\phi$), with the swapping operator $\hat Z_2 = \sum_{n_1,n_2} \ket{n_1,n_2}\bra{n_2,n_1}$ (written in the Fock-state basis of the two individual modes). The swapping operator is a conserved quantity. This is only possible due to the nonlocality of the dissipation \cite{Nigro} and is the origin of \textit{quantum multistability} \cite{VictorAlbert}. Since there are infinitely-many possible initial expectation values of $\hat Z_2$, there are also infinitely-many steady states. This is true for any value of $N$. For a finite $N$, the steady state reads $\hat \rho_\text{ss} = \hat r_{0,1} + c_{0,2} \hat r_{0,2}$, where $\hat r_{0,1}$ is a density matrix while $\hat r_{0,2}$ is not, and the coefficient $c_{0,2} = \Tr[\hat Z_2 \hat \rho(0)]$ depends on the initial condition. This steady state is symmetric, i.e., $\hat U(\phi) \hat \rho_\text{ss} \hat U^\dag (\phi) = \hat \rho_\text{ss}$, irrespective of the initial condition.

{\it Secondly}, the system we consider presents limit cycles \footnote{Note that the limit cycles are not a consequence of having a nontrivial conserved operator ($\hat Z_2$). The same BHD model, but with antisymmetric drive, also exhibits limit cycles and only $\hat{\mathbb 1}$ is conserved.} in the limit $N\to \infty$. These are persistent periodic oscillations in the infinite time limit of the dynamical evolution. Recently, an open quantum system with this property has been termed a {\it boundary time crystal} \cite{Keeling2018}. It is there argued that while the total Hamiltonian (of the boundary, the bulk, and their coupling interaction) is time-translation invariant, the boundary system presents limit cycles, thus breaking the global time-translation symmetry and forming \textit{time crystal} with a nonrigid period that is allowed to change continuously as a function of the system parameters. Our system satisfies the same conditions. The dependence on the period of the coherent drive can be eliminated by a gauge transformation. This means that any emergent periodic response stems from a continuous breaking of time-translation invariance in contrast to the discrete fashion due to the presence of a Floquet-map eigenvalue in the unit circle \cite{Poletti2018, Ueda2018}. The steady state of the BHD, which spontaneously breaks the time-translation invariance of the Liouvillian, assumes the form (for $N \to \infty$)
\begin{equation}\label{equ:rhossLC}
\hat \rho_\text{ss}(t) = \hat r_{0,1} + \sum\limits_{d=2}^{D_0} c_{0,d} \, \hat r_{0,d} + \sum\limits_{d=1}^{D_1} (c_{1,d} e^{i|\lambda_1|t} \hat r_{1,d} + \text{H.c.}),
\end{equation}
where $\L(\hat r_{n, d}) = \lambda_n \hat r_{n,d}$, $c_{n, d}$ are coefficients depending on the initial system density matrix $\hat \rho(0)$, and $\lambda_1$ is a purely imaginary eigenvalue responsible for the formation of a limit cycle \footnote{See the Supplemental Material for further details, which includes Refs. \cite{Jaksch2018, RuboPRB2012, RuboPRL15, GarrahanMetastability, PlenioRevMod, BreuerBook, Marti2017, Ciccarello2017, Choi2017}.}. The upper limits $D_0$ and $D_1$ in the sums of the right-hand side in Eq. \eqref{equ:rhossLC} are unknown. The index $d=1,...,D_n$ labels the degeneracy of the eigenvalue $\lambda_n$. In principle, one could prepare an initial condition such that the coefficients $c_{1,d}$ vanish and the system  reaches a time-independent steady state. For a system prepared in such a manner, an infinitesimal disturbance would lead to $c_{1,d}\neq 0$ with some degeneracy $d$, causing the response to oscillate ceaselessly and forming a time crystal. This could in principle be observed in any time-delayed two-point measurement.

{\it Finally}, the system response is determined by semiclassical complex-amplitude bistability as $N\to \infty$. This is known to occur in the BHD under local dissipation for the same driving configuration as the one we consider here \cite{Wouters2017}. Semiclassical bistability manifests itself as a region in the parameter space where two semiclassical fixed points, with different complex amplitudes, appear. In our model, we will show that two attractors, specifically limit cycles with different periods, coexist in this region.

\section{Results}
In order to assess the system behavior as $N\to \infty$ we resort to the Gross-Pitaevskii (semiclassical) approach, in which the operators in the Heisenberg picture are replaced by their expectation values $\alpha_i = \langle \hat a_i \rangle$. We remark that our continuous swapping symmetry is intrinsically quantum and cannot be captured by the semiclassical picture, where it shows up merely as a discrete swapping symmetry $\alpha_1 \leftrightarrow \alpha_2$. We compare the semiclassical predictions with the calculations in the quantum regime for increasing values of $N$.

In Fig. \ref{fig: figure1}(a) we depict the steady state expectation value of the rescaled total number of bosons $\hat n_\text{tot}/N=(\hat a_1^\dag \hat a_1 + \hat a_2^\dag \hat a_2)/N$ as a function of the rescaled pump amplitude $\tilde F/\gamma = F/(\sqrt{N} \gamma)$. We present in color the quantum results for different values of $N$. The continuous (dashed) black lines indicate stable (unstable) semiclassical fixed points. One can see that as $N$ increases, the occupation number in the quantum regime approaches one of the two semiclassical stable branches. The transition between the two branches becomes increasingly sharper, suggesting the formation of a discontinuous jump, as one would expect from a first-order DPT \cite{Casteels-Fazio-Ciuti_PRA2017, CiutiSpectralTheo}. 

The basin of attraction of the stable fixed points is just the two-dimensional plane given by $\alpha_1=\alpha_2$ in the four-dimensional phase space defined by the real and imaginary parts of $\alpha_1$ and $\alpha_2$. A randomly chosen initial condition generates a trajectory that most certainly avoids the plane, converging to one of the families of limit cycles which revolve around the stable fixed points. The amplitude of the attained limit cycle is determined by the initial difference in the coherent states $\alpha_1(0)-\alpha_2(0)$, although its period is solely dependent on the set of parameters entering in Eq. (\ref{equ: Lindblad equation}), except in the region of bistability, where limit cycles with two different frequencies have different basins of attraction. Interestingly, the limit cycles break the swapping symmetry. They are the so-called \textit{Josephson oscillations}. A typical example of these oscillations is depicted in Fig. \ref{fig: figure1}(b).

In Fig. \ref{fig: figure1}(d), we show the five eigenvalues whose real parts are closer to zero (defining the sequence of the gaps), for increasing values of $N$. We distinguish them using different colors and markers according to the proximity of their real parts to the zero eigenvalue (which is doubly degenerate \textemdash{for} the identity and the swapping operator). We consider only one of the two complex eigenvalues forming a conjugate pair. It is evident that all the gaps are closing with growing $N$. Eigenvalue crossings are also visible, while for moderate pump strengths there is an eigenvalue with an {\it inverted-parabolic shape}, flattening to zero as $N$ increases. This eigenvalue has zero imaginary part, suggesting that it is itself responsible for a first-order DPT in the thermodynamic limit mediated by the presence of bistability \cite{CiutiSpectralTheo}. 

In addition, there are two eigenvalues with a nonzero imaginary part at low and high pump strengths. Their imaginary parts converge very quickly for relatively small $N$ (at $\tilde F/\gamma=1.65$ we have a maximum of $\sim 13$ bosons  for $N=4$). These are the eigenvalues responsible for the limit cycles as $N\to \infty$. To illustrate this point, we show the frequencies of the semiclassical limit cycles (in black squares) in the plot $N=4$. For low pump values, the scaled frequency $\text{Im}[\lambda]/\gamma$ has already reached its $N\to \infty$ value for $N=4$. As we have already anticipated, two limit cycles with different periods coexist in the semiclassical bistable region ($1.02 \lesssim \tilde F/\gamma \lesssim 1.1$). Either one or the other is reached in the long-time limit, depending on the amplitude of the initial coherent state.

The closure of the first gap for all the pump values we have considered suggests a critical slowing down transcending the bistability region, in contrast with local dissipation \cite{Note2}. To show this we employ the time-delayed second-order correlation function \cite{Fink2018} (symmetric with respect to the two modes $i=1,2$) $g^{(2)}(\tau) = \braket{\hat a_i^{\dagger}(0) \hat a_i^{\dagger}(\tau)\hat a_i(\tau) \hat a_i(0)}_{\rm ss}/\displaystyle\braket{\hat a_i^\dag \hat a_i}_{\rm ss}^2 = \Tr[\hat a_i^\dag(0) \hat a_i(0) e^{\tau\L}(\hat \rho')]/\braket{\hat a_i^\dag \hat a_i}_{\rm ss}$, where $\hat \rho' = \hat a_i \hat \rho_\text{ss} \hat a_i^\dag/\Tr[\hat a_i^\dag \hat a_i \hat \rho_\text{ss}]$ is the state of the system after an initial detection of the mode $i$ and is evolved during a time $\tau$ before making a second measurement. The subscript ${\rm ss}$ refers to the steady state. After a sufficiently long time, if the steady state is unique, $g^{(2)}(\tau)$ will relax to the value of $1$, due to its normalization. In Fig. \ref{fig: figure1}(c) we plot $g^{(2)}(\tau)$ for three pump values. One can observe that the correlations take a long time to decay for $\tilde F/\gamma=0.5$, far below the bistability region. Furthermore, the function does not always converge to $1$. This is a clear manifestation of the infinite number of steady states: The initial detection changes the value of the conserved quantity $\langle \hat Z_2 \rangle$, so the system must relax to a different steady state.

\begin{figure}
\includegraphics[width=1\linewidth]{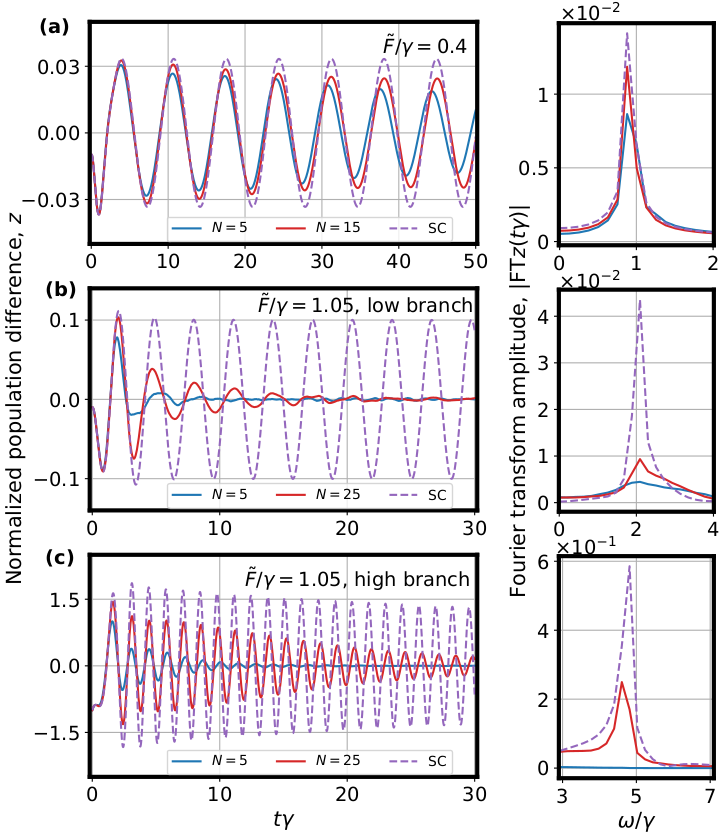} 
\caption{Emergence of limit-cycle oscillations. Normalized population difference $z = \langle \hat a_1^\dag \hat a_1 - \hat a_2^\dag \hat a_2 \rangle/N$ as a function of time and the respective Fourier transform for two different pump values: $\tilde F/\gamma=0.4$ in (a) and $\tilde F/\gamma =1.05$ in (b) and (c). The initial condition is a scaled coherent state $\ket{\alpha_1=0, \alpha_2=0.1 \sqrt{N}}$ in (a) and (b) and $\ket{\alpha_1=0, \alpha_2= \sqrt{N}}$ in (c). In the bistability region ($\tilde F/\gamma =1.05$) these initial states lead to the lower and upper semiclassical branches [see Fig. \ref{fig: figure1}(a)], respectively. For finite $N$, the results in (a) are obtained by solving Eq. (\ref{equ: Lindblad equation}), while in (b) and (c) by averaging over $5000$ quantum trajectories. The Fock space per site is truncated at 30 bosons for $N=15$ in (a) and at 90 for $N=25$ in (b) and (c). The rest of the parameters are the same as in Fig. \ref{fig: figure1}.}
\label{fig: emergence}
\end{figure}

Figure \ref{fig: emergence} clearly depicts the emergence of the periodic oscillations as $N\to \infty$. In frames (b) and (c), we can observe frequency bistability, starting from two different initial conditions. Here we are in the semiclassical bistability region with finite $N$, where the oscillations are damped. Moreover, we show in Ref. \cite{Note2} that single trajectories Fourier transform as well to a neat frequency peak matching that of the semiclassical response (apart from a small deviation due to stochastic jumps).

In the bistability region, a single quantum trajectory evidences switching between the dim and bright metastable states \cite{Carmichael2015}. This is accompanied by a frequency switch for the oscillations between the two (frequency) branches drawn in Fig. \ref{fig: figure1}(c) for $N=4$. However, for large but finite $N$, the inverse of the switching rate is larger than the lifetime of the oscillations, which means that on average one obtains an oscillation frequency corresponding to either the lower or the upper branches of Fig. \ref{fig: figure1}(c), depending on the initial condition [see Figs. \ref{fig: emergence}(b) and \ref{fig: emergence}(c)].

In Fig. \ref{fig: emergence}(c), one can notice a mismatch between the frequency peaks in the semiclassical and the quantum regimes for $N=25$. There are two reasons for this: (i) the semiclassical evolution remains transient and the frequency has not yet attained its limit-cycle value (which depends nonlinearly on the mode population and therefore decreases in the long-time limit), (ii) the imaginary part of the Liouvillian eigenvalue responsible for the oscillations has not yet converged to its $N\to \infty$ value.
 
We now propose a mechanism accounting for the observed periodic oscillations. When $\tilde U=0$, bonding and antibonding modes are decoupled, and the antibonding mode evolves coherently giving trivial periodic oscillations in both the quantum and semiclassical regimes. When $\tilde U\neq 0$, the interaction term becomes very inefficient in coupling both modes when the antibonding-mode population is sufficiently small. The emergence of the limit cycles is then due to an effective decoupling between the two modes (see also Ref. \cite{Note2} for further details).

\section{Discussion}
Nonlocal dissipation for open quantum systems without detailed balance has been considered in the context of noiseless subsystems or decoherence-free subspaces \cite{Zanardi97,*Zanardi98, Lidar98}, in cold atoms or many-body spin systems \cite{Keeling2018, James93, Olmos14, Cooper18}, and in a recent proposal of a Floquet time crystal \cite{Poletti2018}. In general, that type of dissipation is either engineered or appears as a consequence of photon-mediated long range interactions. The specific form of Eq. (\ref{equ: Lindblad equation}) can be derived \cite{Note2} from first principles if one considers that both modes in our system have the exact same coupling with a single bosonic bath. This approach is similar to Ref. \cite{Zanardi97,*Zanardi98}.

In practice, the assumption of identical coupling between the two modes and the environment is idealistic. While the coupling strength between the two system modes and the reservoir modes (with frequencies in the vicinity of the driving frequency) could be equal, in general the relative phases will be different \cite{Note2}. This implies a mixture of bonding and antibonding dissipation and, in turn, the mixture of local and nonlocal dissipation following a change of basis. An important question, then, concerns the implications of a broken symmetry with respect to the swapping operation in the presence of a small yet non-negligible amount of dissipation for the antibonding mode. This perturbation has the consequence of lifting the degeneracy of the zero eigenvalue in the Liouvillian spectrum and also destroying the limit cycles. However, as we show in Ref. \cite{Note2}, periodic oscillations persist for long times, and the time-crystalline period is robust.

The BHD has already been studied in the laboratory \cite{Rodriguez2017}. The type of dissipation in our model could be realized in a circuit QED experiment coupling two degenerate modes to a microwave resonator at a single spatial location, where the wave function overlap between the resonator mode and both system modes is identical. Alternatively, it could be realized with polaritons in two coupled semiconductor micropillars if the bonding and antibonding mode linewidths differ by two orders of magnitude or more. This could be achieved by etching the sample in such a way that the nonradiative losses \cite{Alberto2018} of the bonding mode make the dominant part of the total dissipation.

\begin{acknowledgments}
We are thankful to the developers of the Quantum Toolbox in Python (QuTiP) \cite{Qutip1, Qutip2}, as we have used it for most of our calculations in the quantum regime. We thank A. Amo, E. Ginossar, and M. Stern for helpful discussions, and J. Keeling for the careful reading of the manuscript. C.L. gratefully acknowledges the financial support of CONICYT through Becas Chile 2017, Contract No. 72180352. M.H.S. gratefully acknowledges financial support from QuantERA InterPol and EPSRC (Grant No. EP/R04399X/1 and No. EP/K003623/2). 
\end{acknowledgments}

\bibliography{PRBSubmission}

\end{document}


\def\mean#1{\left< #1 \right>}

\title{Supplementary Material for \protect\\ \textit{Driven Bose-Hubbard dimer under nonlocal dissipation: A bistable time crystal}}

\author{C. Lled\'{o}}
\email[electronic address: ]{c.lledo.17@ucl.ac.uk}
\affiliation{Department of Physics and Astronomy, University College London,
Gower Street, London, WC1E 6BT, United Kingdom}
\author{Th. K. Mavrogordatos}
\affiliation{Department of Physics, Stockholm University, SE-106 91, Stockholm, Sweden}
\author{M. H. Szyma\'{n}ska}
\affiliation{Department of Physics and Astronomy, University College London,
Gower Street, London, WC1E 6BT, United Kingdom}

\maketitle

In the Supplementary Material we provide further details concerning the role of conserved quantities and symmetries in the Bose-Hubbard dimer (BHD), viewed explicitly as an open quantum system. We give further insights on the importance of the conserved swapping operator ($\hat{Z}_2$) in the quantum regime when we approach the first-order \textit{Dissipative Phase Transition} (DPT) of amplitude bistability and its associated metastability. We visualize the quantum fluctuations through the Q-function quasiprobability distribution and quantum-jump trajectories for pump values  around the bistability threshold, as the $N\to\infty$ thermodynamic limit is approached. We also provide further links to the semiclassical treatment and explain the effective decoupling (between the bonding and antibonding modes) mechanism responsible of the time-crystal behaviour. For all the results in the main text as well as in this analysis, we compare with what is expected from the BHD under {\it local dissipation}. Moreover, we show how the Lindblad Master Equations (LMEs) with local and nonlocal dissipation can be derived from a total Hamiltonian that includes the environment and its coupling to the BHD. Finally, we explore the possibility of an asymmetric system-environment coupling that destroys the continuous swapping symmetry and the limit cycles in the nonlocal dissipation case. We show that, in spite of this, the Josephson-like oscillations remain long lived, with a robust time-crystalline period.

\tableofcontents

\section{Symmetries and Conserved quantities for a Markovian open quantum system}
\label{sec: symmetris and conserved quantities}

In order to understand the occurrence of dissipative phase transitions, quantum multistability, and quantum limit cycles, it is crucial to investigate the role of symmetries and conserved quantities in open quantum systems, and how these are related to the spectrum of the Liouvillian. We introduce here some analytical properties which we invoke in the main text to appreciate the peculiarities about nonlocal dissipation as opposed to local dissipation.

Let us first introduce the dual super-operator of the Liouvillian. For a general $\L(\cdot) = -i[\hat{\mathcal H}, \cdot] + \sum_{k} \gamma_k (\hat L_k \cdot \hat L_k^\dag - (1/2)\{ \hat L_k^\dag \hat L_k, \cdot \} )$,  its dual
\begin{equation}
\L^*(\cdot) = i[\hat H, \cdot] + \sum_{k} \gamma_k \left(\hat L_k^\dag \cdot \hat L_k -\frac{1}{2}\{\hat L_k^\dag L_k, \cdot \} \right)
\end{equation}
is the generator of the evolution of system operators, i.e., $\partial_t \hat A = \L^*(\hat A)$ for any system operator $\hat A$. It is easy then to show by direct inspection that when the Hamiltonian and every Lindblad operator commutes with an operator $\hat O$, i.e.,
\begin{equation}
 [\hat H, \hat O] = [\hat L_k, \hat O]=0 \quad \forall k,
\end{equation}
then $\hat O$ is a conserved quantity, with $\partial_t \hat O = \L^*(\hat O) = 0$, and there is a continuous symmetry $\hat U(\phi) = \exp(i\phi \hat O)$ (for real $\phi$) such that $\L(\hat U(\phi) \hat \rho \hat U(\phi)^\dag) = \hat U(\phi) \mathcal L(\hat \rho) \hat U(\phi)^\dag$ for any $\hat\rho$. This is a one way implication only ($\Rightarrow$). Examples which show that this is not an equivalence and that a continuous symmetry does not necessarily have a conserved quantity and \textit{vice versa} can be found in \cite{VictorAlbert}. 

In general, the Liouvillian $\mathcal L$ can be subject to a spectral decomposition. (An equivalent formulation of these statements in the so-called \emph{Liouville space} can be found in \cite{VictorAlbert}). Let us consider the set of operators $\{\hat r_{n,d_n},\hat l_{n,d_n}\}$ which are eigenoperators of the Liouvillian $\L$ and its dual $\L^*$:
\begin{equation}
 \L(\hat r_{n,d_n}) = \lambda_n \hat r_{n,d_n}, \quad \L^*(\hat l_{n,d_n}) = \lambda_{n}^{*} \hat l_{n,d_n},
\end{equation}
and are orthonormal in the internal \emph{Hilbert-Schmidt} dot product
\begin{equation}
\Tr[\hat l_{j,d_j}^\dag \hat r_{k, d_k}] = \delta_{jk} \delta_{d_j,d_k},
\end{equation}
where the sub-index $d_n=1,2,3,\dots$ labels the degeneracy of the eigenvalue $\lambda_n$. The set of operators $\{\hat r_{n,d_n}, \hat l_{n,d_n}\}$ form a basis of bounded operators in the system Hilbert space, thus the solution of the Lindblad equation $\hat \rho(t) = e^{t\L}\hat \rho(0)$ can be expanded as
\begin{equation} \label{equ: Lindblad evolution expansion}
 \hat \rho(t) = \sum_{n,d_n} c_{n,d_n} e^{t\L} \hat r_{n,d_n} = \sum_{n,d_n} c_{n,d_n} e^{t\lambda_n} \hat r_{n,d_n},
\end{equation}
with complex coefficients
\begin{equation}
 c_{n,d_n} = \Tr[\hat l_{n,d_n}^\dag \hat \rho(0)],
\end{equation}
depending on the initial condition. Note that $\hat r_{n,d_n}$ and $\hat l_{n,d_n}$ are not necessarily density matrices, in the sense that they might not be semidefinite positive, hermitian, or have a unit trace.

Since in general the Liouvillian is not the same as its dual, the eigenvalues $\lambda_n$ are complex. It becomes clear from Eq. (\ref{equ: Lindblad evolution expansion}) that in order to describe a physical evolution (a {\it CPTP map}) they must have nonnegative real parts, $\text{Re}[\lambda_n]\le 0$, or else the density matrix would grow exponentially in time. This means that their real parts are decay rates while their imaginary parts give rise to oscillations. We can order them such that $0\geq\text{Re}(\lambda_0)\geq \text{Re}(\lambda_1) \geq \text{Re}(\lambda_2) \geq \ldots$. There is always an eigenvalue equal to zero ($\lambda_0=0$) since the identity satisfies $\mathcal L^*(\hat{\mathbb{1}}) = 0$. This is due to the trace-preserving condition of the evolution. In other words, the identity is trivially conserved. If the identity is the only conserved operator and there is no other purely imaginary eigenvalue, then the steady state of Eq.(\ref{equ: Lindblad evolution expansion}) assumes the form
\begin{equation}
 \hat \rho_\text{ss} = \lim_{t\to\infty} \hat \rho(t) = c_{0,1} \hat r_{0,1} = \hat r_{0,1},
\end{equation}
since $c_{0,1} = \Tr[\hat{\mathbb{1}} \hat \rho(0)] = \Tr[\hat \rho(0)] = 1$. This steady state is unique and carries no information about the initial condition. We note that, as proved in \cite{Nigro}, for any finite Hilbert space cutoff, all systems with local bosonic Lindblad operators of the form $\hat a_i$ acting on the mode (or site) $i$, possess a unique steady state or, equivalently, cannot have any nontrivial conserved operator. This is the case of the local dissipation for the BHD that is most commonly considered in the literature (see e.g. \cite{CaoPRA,Casteels-Ciuti_PRA2017}).

If there is, however, a nontrivial conserved operator $\hat O$, then the steady state will take the form
\begin{equation} \label{equ: steady state with one conserved quantity}
 \hat \rho_\text{ss} = \hat r_{0,1} + c_{0,2} \hat r_{0,2},
\end{equation}
where the coefficient $c_{0,2} = \Tr[\hat O^\dag \hat \rho(0)]$ preserves some information of the initial condition. The infinite number of possible steady states is a direct consequence of the presence of a nontrivial conserved quantity. Note that the following orthonormality  conditions hold:
\begin{equation}
 \begin{split}
  1 &= \Tr[\hat{\mathbb 1} \hat r_{0,1}] = \Tr[\hat r_{0,1}], \\
  0 &= \Tr[\hat{\mathbb 1} \hat r_{0,2}] = \Tr[\hat r_{0,2}], \\
  0 &= \Tr[\hat O^\dag \hat r_{0,1}], \\
  1 &= \Tr[\hat O^\dag \hat r_{0,2}].
 \end{split}
\end{equation}
In particular, an initial condition can be chosen such that $c_{0,2}=0$, thus $\hat r_{0,1}$ is always a density matrix.

It is possible to have a purely imaginary eigenvalue (together with its complex conjugate), for instance $\lambda_1 = i \text{Im}[\lambda_1]$ (and $\lambda_1^{*} = -i \text{Im}[\lambda_1]$), which would give rise to a quantum limit cycle \cite{[{A time crystal has a limit cycle in the $N\to\infty$ limit, spontaneously breaking the time-translation symmetry. See: }]Keeling2018,*Ueda2018,*Poletti2018}. If there are no conserved quantities (besides the identity), in the long-time limit, the steady state will have the oscillatory form
\begin{equation} \label{equ: quantum limit cycle steady state}
 \hat \rho_\text{ss}(t) = \hat r_{0,1} + c_{1,1} e^{\lambda_1\,t} \hat r_{1,1} + c_{1,1}^* e^{\lambda_1^{*}\,t} \hat r_{1,1}^\dag, 
\end{equation} 
with $c_{1,1} = \Tr[\hat l_{1,1}^\dag \hat \rho(0)]$.

 
\subsection{Symmetries and conserved quantities for the BHD}

The local dissipation Liouvillian $\L_\text{loc}(\cdot) = -i[\hat{\mathcal H}, \cdot] + \gamma \mathcal D[\hat a_1](\cdot) + \gamma \mathcal D[\hat a_2](\cdot)$, with the Hamiltonian $\hat{\mathcal H}$ of the BHD given in Eq.(2) in the main text, has a discrete $Z_2$ symmetry, $\hat a_1 \leftrightarrow \hat a_2$, which can be readily checked by inspection. Formally, this is given by the unitary swapping operator $\hat Z_2 = \sum\limits_{n_1, n_2} \ket{n_1, n_2} \bra{n_2, n_1}$, written in the Fock-state basis of the two modes (with $n_1, n_2$ their occupation number). Then, using the fact that $\hat Z_2 \hat a_1 \hat Z_2 = \hat a_2$ and $\hat Z_2 \hat a_2 \hat Z_2 = \hat a_1$ (note that $\hat Z_2^{-1} = \hat Z_2$), one can check that
\begin{equation}
 \mathcal L_\text{loc}(\hat Z_2 \hat\rho \hat Z_2) = \hat Z_2 \mathcal L_\text{loc} (\hat \rho) \hat Z_2.
\end{equation}
Since $\partial_t \hat \rho  = \L_\text{loc}(\hat \rho)$ has a unique steady state \cite{Nigro}, it follows that $\hat \rho_\text{ss}^\text{loc} = \hat Z_2 \hat \rho_\text{ss}^\text{loc} \hat Z_2$; hence, if one traces out any of the two modes the density matrix remains the same:
\begin{equation} \label{equ: discrete symmetry Z_2}
\hat \rho_\text{ss,1}^\text{loc} = \Tr_2[\hat \rho_\text{ss}^\text{loc}] = \Tr_1[\hat \rho_\text{ss}^\text{loc}] = \hat \rho_\text{ss,2}^\text{loc}.
\end{equation}
Note that this is only a discrete symmetry and no continuous symmetry can be generated by $\hat Z_2$ since it does not commute with both Lindblad operators $\hat a_1$ and $\hat a_2$ individually.

For the nonlocal dissipation case, since $\hat Z_2$ commutes with the Hamiltonian and with the unique Lindblad operator ($\hat a_1 + \hat a_2$), the symmetry is continuous and $\hat Z_2$ is conserved, thus,
\begin{equation}
 \mathcal L( \hat U(\phi) \hat \rho \hat U(\phi)^\dag) = \hat U(\phi) \mathcal L(\hat \rho) \hat U(\phi)^\dag \quad \text{and} \quad \partial_t \hat Z_2=0,
\end{equation}
with $\hat U(\phi) = e^{i\phi\hat Z_2} = \cos(\phi) \hat{\mathbb 1} + i\sin(\phi) \hat Z_2$. As stated in the main text, the steady state $\hat \rho_\text{ss}$ is always symmetric independently of the initial condition for any finite system size $N$. This is not necessarily  true when the $\lambda_0=0$ eigenvalue becomes degenerated in the thermodynamic limit, where a spontaneous symmetry breaking can take place.

\subsection{Symmetries and conserved quantities in the absence of the nonlinearity} \label{sec: symmetries and conserved w/o non-linearity}

An alternative representation that is helpful to understand the difference between the local and nonlocal dissipation approaches is obtained by rewriting the Liouvillians $\L$ and $\L_\text{loc}$ in terms of the bonding and anti-bonding bosonic operators $\hat a_{B, A} = (\hat a_1 \pm \hat a_2)/\sqrt{2}$, respectively. The Hamiltonian of Eq. (1) in the main text becomes
\begin{equation} \label{equ: bonding and antibonding Hamiltonian}
\begin{split}
 \hat{\mathcal H} = &(-\Delta-J) \hat a_B^\dag \hat a_B + (-\Delta + J) \hat a_{A}^\dag \hat a_{A} + \sqrt{2}F(\hat a_{B}^\dag + \hat a_{B}) \\
+&\frac{U}{4}\left[ \sum\limits_{k=B,A}(\hat a^\dag_k \hat a^\dag_k \hat a_k \hat a_k) + 2\hat a_B^\dag \hat a_B^\dag \hat a_A \hat a_A + 2\hat a_B \hat a_B \hat a_A^\dag \hat a_A^\dag \right. \\
&+\ 8 \hat a_B^\dag \hat a_B \hat a_A^\dag \hat a_A \Bigg].
\end{split}
\end{equation}
The local dissipator $\mathcal D_\text{loc}(\cdot) = \gamma \mathcal D[\hat a_1](\cdot) + \gamma \mathcal D[\hat a_2](\cdot)$ is rewritten as 
\begin{equation}
\mathcal D_\text{loc}(\cdot) = \gamma \mathcal D[\hat a_B](\cdot) + \gamma \mathcal D[\hat a_A](\cdot),
\end{equation}
while the nonlocal dissipator $\mathcal D(\cdot) = \gamma \mathcal D[\hat a_1 + \hat a_2](\cdot)$ is
\begin{equation}
 \mathcal D(\cdot) = 2\gamma \mathcal D[\hat a_B](\cdot).
\end{equation}

We note that in the Hamiltonian of Eq. \eqref{equ: bonding and antibonding Hamiltonian} the driving changes only the occupation of the bonding mode, while the nonlinear terms couple the bonding and antibonding modes. This means that if we remove the nonlinearity ($U=0$), we have that the Hamiltonian conserves the antibonding mode occupation number $[\hat{\mathcal H}_\text{linear}, \hat n_{A}]=0$ where $\hat n_{A} = \hat a_A^\dag \hat a_A$. While both modes are dissipated in the local approach, only the bonding mode is dissipated in the nonlocal approach, which means that $\hat n_A$ is a conserved operator when the dissipation is nonlocal, $\partial_t \hat n_A = \mathcal L_\text{linear}^*(\hat n_A) = 0$, and is associated with a continuous antibonding rotation symmetry given by the unitary transformation $\hat R = e^{i\theta \hat n_{A}}$ (for real $\theta$). For the local dissipation approach, the antibonding mode is dissipated such that $\hat n_A$ is not a conserved quantity, even though $\hat R$ defines a continuous symmetry, i.e., $\L_\text{loc}(\hat R \hat \rho \hat R^\dag) = \hat R \L_\text{loc}(\hat \rho) \hat R^\dag$. Since $\hat{\mathbb{1}}$, $\hat Z_2$, and $\hat n_A$ are linearly independent, we conclude that in the absence of the nonlinearity there are three conserved quantities in the nonlocal approach when driving is symmetric ($F_1=F_2$).

Note that the bonding and antibonding modes are completely decoupled in the dynamics if $U=0$. Under local dissipation the antibonding mode is damped due to the dissipation, whereas under nonlocal dissipation the antibonding mode evolves coherently (i.e. without dissipation) as we mention in the main text. This means there is a purely imaginary Liouvillian eigenvalue ($\lambda = i(J-\Delta)$) responsible for trivial coherent oscillations. As we discuss later, this decoupling is essential to our understanding of the emergence of limit cycles in the BHD as an effective decoupling between the bonding and antibonding modes \cite{Jaksch2018}.

In principle, there could be other conserved quantities in the local or nonlocal dissipation cases. However, by numerically calculating the degeneracy of the zero eigenvalue, one can check that for local dissipation there is only one conserved quantity (the identity) while for the nonlocal case, one has two conserved quantities ($\hat{\mathbb{1}}$ and $\hat Z_2$) in the presence of the nonlinearity, and an additional one ($\hat n_A$) when the nonlinearity is removed. There could be other continuous symmetries we are not aware of, but these would not be associated with conserved quantities.

We also remark that the zero eigenvalue corresponding to the conserved operator $\hat n_A$ in the nonlocal dissipation case might be absent in a numerically-obtained spectral decomposition of $\L_\text{linear}$ because $\hat n_{A}$ does not commute with $\hat{\mathcal H}$ or $\hat a_B$ for a finite Hilbert-space cutoff. This is not the case for $\hat Z_2$, which commutes with both operators for any cutoff.

\begin{figure*} 
\centering
\includegraphics[width=1\textwidth]{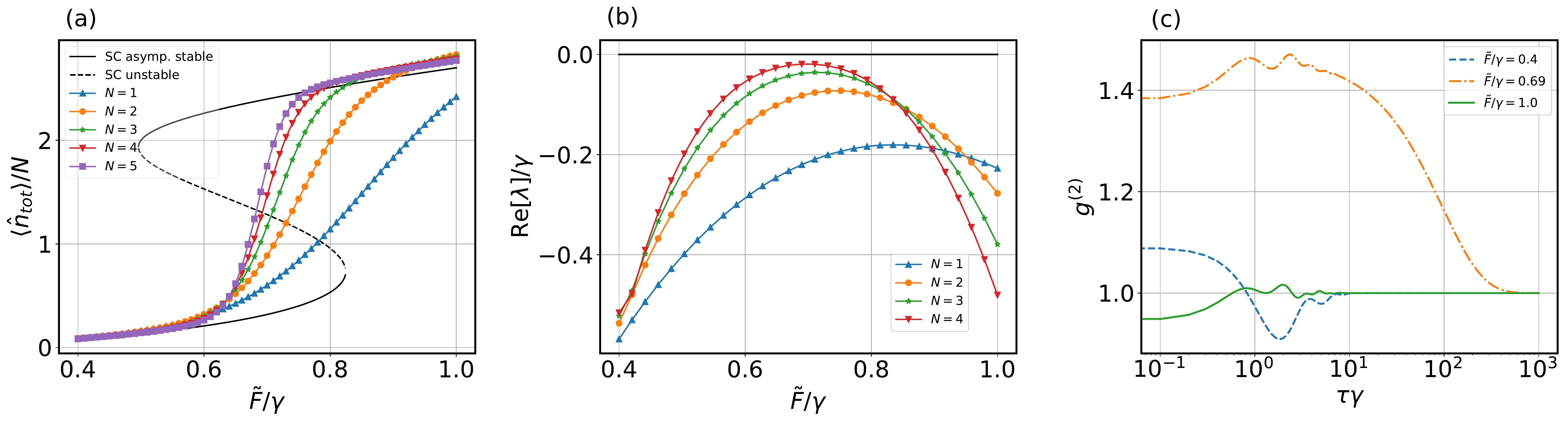}
\caption{\textbf{Bistability for local dissipation.} Frames (a) and (b) show the total number of bosons $\langle \hat n_\text{tot}\rangle/N$ in the steady state and the real part of the Liouvillian gap, $\text{Re}[\lambda]/\gamma$, respectively, as a function of the scaled drive $\tilde F/\gamma = N^{-1/2} F/\gamma$ for different values of $N$. In (a) the black solid and dashed lines, denoted in the legend by {\it SC}, depict semiclassical calculations. Frame (c) shows the time-delayed second-order coherence function $g^{(2)}(\tau \gamma)$ for three different pump values and $N=5$. The parameters used read: $\Delta=J= \tilde U= \gamma$ and $U=\tilde U/N$.}
\label{fig: Bistability plots}
\end{figure*}

\section{Dissipative phase transitions}
\label{sec: DPTs}

It is a well-known fact that, due to the nonlinearity, the semiclassical regime of the BHD with local dissipation can present bistability (for symmetric driving) \cite{Wouters2017} or undergo a symmetry breaking (for anti-symmetric driving)\cite{Casteels-Ciuti_PRA2017} in the steady state, as a scaled system parameter is varied.
These two phenomena are associated with dissipative phase transitions. In analogy with a \textit{closed} quantum system, where the excitation gap of the Hamiltonian in the thermodynamic limit may close when a parameter is varied, in an \textit{open} quantum system the spectrum gap of the Liouvillian ($\mathcal L$) may also close for an appropriately-defined thermodynamic limit. In the former case, the transition is termed  \textit{quantum phase transition}, while in the latter,  \textit{dissipative phase transition} (DPT) \cite{Kessler2012, CiutiSpectralTheo}.

In simple words, if the gap in the Liouvillian spectrum, i.e., the eigenvalue $\lambda \equiv \lambda_1$, tends towards zero in the thermodynamic limit for either a specific value of a varying parameter ($\xi=\xi_c$) or in a whole region of $\xi \in \Xi$, then a \textit{first-order} or \textit{second-order} DPT may occur. Since $\text{Re}[\lambda]$ is the smallest decay rate, the system response exhibits a \textit{critical slowing down} as that rate approaches zero, which means that it takes a very long time ($\sim \text{Re}[\lambda]^{-1}$) to reach the steady state. This behaviour has been observed experimentally in photon time-delayed correlation measurements in single cavity polaritons \cite{Fink2018}. The difference between the DPTs of first and second order is that in the former there is a nonanalyticity at the single critical point $\xi_c$ due to a level crossing, while in the latter the eigenoperator associated with the eigenvalue of the closing gap renders the steady-state subspace degenerated \cite{CiutiSpectralTheo}, hence subject to energy-costless excitations that break the symmetry (i.e., Goldstone modes).

In the following subsection, we spend some time on the semiclassical results derived from our model, and explain the relevance to the quantum picture.

\begin{figure} 
\centering
\includegraphics[width=0.4\textwidth]{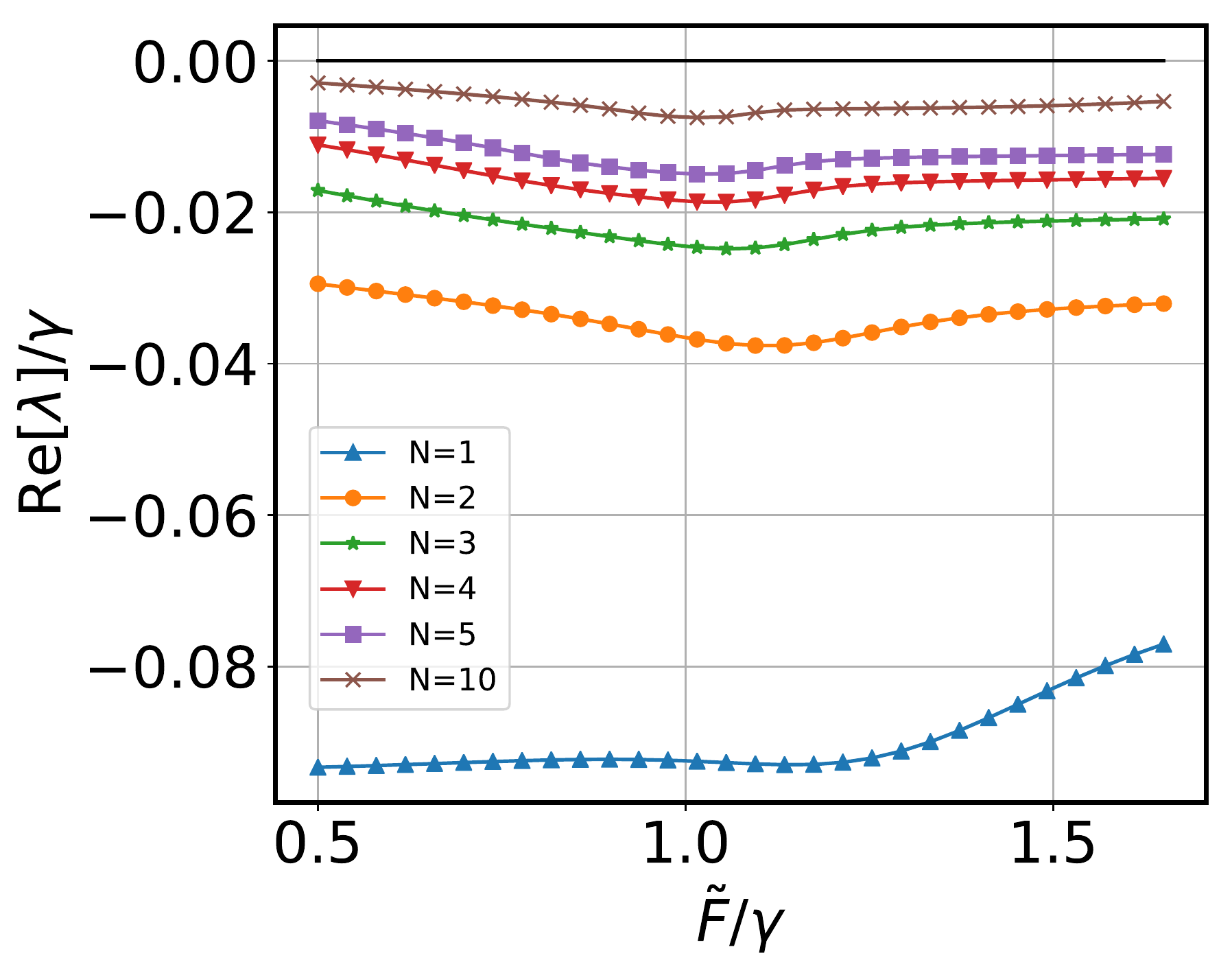}
\caption{\textbf{First Liouvillian gap for nonlocal dissipation} as a function of the scaled drive $\tilde F/\gamma = N^{-1/2}F/\gamma$ for $N=[1-5, 10]$. The parameters used read: $\Delta/\gamma = 0.7$, $J=1.5$, $\tilde U/\gamma=1.0$ and $U=\tilde U/N$.}
\label{fig: non-local Liovillian gap}
\end{figure}

\subsection{The Gross-Pitaevskii semiclassical equations}

Starting from the LME for local or nonlocal dissipation, a semiclassical approximation can be obtained by computing $ \Tr[\hat a_i \partial_t\hat \rho]$ assuming that the state of the system is uncorrelated and coherent, with a density matrix of the form $\hat \rho = \ket{\alpha_1, \alpha_2}\bra{\alpha_1,\alpha_2}$. One then obtains two coupled nonlinear differential equations for the complex-field amplitudes:
\begin{equation} \label{equ: Gross-Pitaevskii equations}
\begin{split}
i \partial_t \alpha_1 &= (-\Delta - i\gamma/2 + 2U|\alpha_1|^2)\alpha_1 - (J+i\epsilon \gamma/2)\alpha_2 + F, \\
i \partial_t \alpha_2 &= (-\Delta - i\gamma/2 + 2U|\alpha_2|^2)\alpha_2 - (J+i\epsilon \gamma/2)\alpha_1 + F,
\end{split}
\end{equation}
where $\epsilon=0$ ($\epsilon=1$) is for local (nonlocal) dissipation. With the scaling introduced in the main text, the coupled equations (\ref{equ: Gross-Pitaevskii equations}) become
\begin{equation} \label{equ: Gross-Pitaevskii equations rescaled}
\begin{split}
i \partial_t \tilde \alpha_1 &= (-\Delta - i\gamma/2 + 2 \tilde U|\tilde\alpha_1|^2) \tilde \alpha_1 -(J + i\epsilon \gamma/2) \tilde \alpha_2 + \tilde F, \\
i \partial_t  \tilde \alpha_2 &= (-\Delta -i\gamma/2 + 2 \tilde U| \tilde \alpha_2|^2) \tilde \alpha_2 - (J + i\epsilon \gamma/2) \tilde \alpha_1 + \tilde F.
\end{split}
\end{equation}
Here we have rescaled the complex amplitudes as $\alpha_i = \sqrt{N} \tilde \alpha_i$. These equations are valir for any $N$. In particular, as $N\to \infty$ the number of particles diverges as $|\alpha_i|^2 = |\tilde \alpha_i|^2 N \to \infty$. We remark that $N$ is an unphysical parameter conveniently introduced to define a weak-coupling thermodynamic limit, and is not related to the total number of particles itself.

We note that in the so-called \textit{weak-lasing} regime of exciton-polaritons \cite{RuboPRB2012}, equations similar to Eq.(\ref{equ: Gross-Pitaevskii equations}) are used, where two coupled Bose-Einstein condensation centers are described by their mean-field order parameters, under collective dissipation. It has been shown that such a model has limit cycles \cite{RuboPRL15}. The main difference with our semiclassical model is that in our case the pump is coherent, while weak-lasing occurs under incoherent pump.

The semiclassical equations (\ref{equ: Gross-Pitaevskii equations rescaled}) are accompanied by their complex conjugates. We therefore introduce the compact vector notation
\begin{equation} \label{equ: Gross-Pitaevski vectorial notation}
\partial_t  \tilde{\bm{\alpha}}(t) = \bm{G}(\tilde{\bm{\alpha}}), \qquad \tilde{\bm{\alpha}}(t) = \begin{pmatrix}
\tilde \alpha_1(t) \\ \tilde \alpha_1^*(t) \\ \tilde \alpha_2(t) \\ \tilde \alpha_2^*(t)                                                           \end{pmatrix},
\end{equation}
where $\bm{G}$ is the vector field that depends nonlinearly on the $\tilde \alpha_i$. We investigate numerically the stationary solutions $\tilde{\bm{\alpha}}(t) = \tilde{\bm{\alpha}}$ of the system defined by Eq. (\ref{equ: Gross-Pitaevski vectorial notation}). We then analyze their stability by expanding the vector $\tilde{\bm{\alpha}}$ around the stationary solutions up to first order in the fluctuations, obtaining a linear first-order differential equation for the fluctuation vector $\bm{\delta}(t) = (\delta_1(t), \delta_1^*(t), \delta_2(t), \delta_2^*(t))^T$:
\begin{equation}
 \partial_t \bm{\delta}(t) = \bm M \bm \delta(t),
\end{equation}
where $\bm M = \partial \bm G/\partial \tilde{\bm \alpha}$ is the Jacobian matrix evaluated for the steady-state solution under consideration.

The Jacobian $4\times 4$ matrix $\bm M$ generally has complex eigenvalues $\{w_j\}_{j=1}^4$. If any of the four eigenvalues has a positive real part, the solution is \textit{unstable}, as the fluctuations will grow exponentially with time. If all of them have nonpositive real parts, $(\text{Re}\,w_j \leq 0)$ , the solution is \textit{stable}, while if they all have negative real part ($\text{Re} \, w_j < 0$), the solution is called \textit{asymptotically stable}. The difference between these last two is that only in the latter every fluctuation due to a perturbation decays in time. For some parameters values, we have also integrated Eq. (\ref{equ: Gross-Pitaevski vectorial notation}) in time to probe the presence of a limit cycle (i.e., a periodic orbit in the phase portrait reached in the long-time limit).

\section{Local vs nonlocal dissipation}

In this Section, we discuss the differences arising between the semiclassical and the quantum description for local and nonlocal dissipation.

The BHD with local dissipation exhibits semiclassical bistability. In Fig. \ref{fig: Bistability plots}(a) we show the rescaled total number of excitations in the system, $\langle \hat n_\text{tot}\rangle/N =( \langle \hat a_1^\dag \hat a_1 \rangle + \langle \hat a_2^\dag \hat a_2 \rangle)/N$ (which in the semiclassical approximation corresponds to $|\tilde \alpha_1|^2 + |\tilde \alpha_2|^2$) as a function of the rescaled pump amplitude $\tilde F/\gamma = F/(\sqrt{N}\gamma)$. The black solid lines are two branches of asymptotically-stable fixed points while the black dashed line is an unstable branch. There is a pump amplitude window in which two asymptotically stable solutions are allowed. For a given initial condition the response reach either one or the other, which is the cause of hysteresis cycles if one is to ramp the pump up and down as in \cite{Rodriguez2017}. The basin of attraction of these asymptotically stable solutions (i.e., the set of initial-condition points which asymptotically converge with time into the given fixed point) appears to be the complete two-dimensional complex space $\mathbb{C}^2$ (or four-dimensional real space), as all the initial conditions we have tried converge to the corresponding fixed point, as shown in Fig. \ref{fig: Bistability plots}(a) (or, in the region of bistability, to one of the two fixed points). We also show the total number of bosons in the quantum regime for different values of $N$. Similarly to the case of nonlocal dissipation (see Fig. 1(a) in the main text), one can see that as $N$ increases, the occupation number approaches one of the two semiclassical stable branches.

Besides the absence of limit cycles in the local dissipation approach, another important difference in this approach is that there is only a first-order DPT and no symmetry breaking. In Fig. \ref{fig: Bistability plots}(b) we show the Liouvillian ($\L_\text{loc}$) gap as a function of the rescaled pump amplitude for different values of $N$ (its imaginary part is zero for all pump values we consider). This should be contrasted with the first gap of the Liouvillian in the nonlocal approach shown in Fig. 1(c) in the main text, which we have also plotted here in Fig. \ref{fig: non-local Liovillian gap} for higher values of $N$. In the local case, it only closes asymptotically at the bistability threshold. The first gap of $\L_\text{loc}$ has a negative-parabola shape, similar to the eigenvalue at medium pump shown in Fig. 1(c) in the main text for nonlocal dissipation, which we ascribe responsible of the bistability. We remark that the second gap of $\L_\text{loc}$ (not shown) approaches a finite real value as $N$ increases, thus no limit cycles can be formed. In Fig. \ref{fig: Bistability plots}(c) we show the time-delayed second-order coherence function $g^{(2)}(\tau\gamma)$ (defined in the main text) for local dissipation. We consider pump values before, inside, and above the bistability region. It is clear that the critical slowing down only occurs at the bistability threshold (orange dashed-line), as expected from the smallness of the Liouvillian gap in this region (see Fig. \ref{fig: Bistability plots}(b)). A clear manifestation that in this case the steady state is unique is that, for the three pump values, at long times the system relaxes to a steady state which is the same to the starting one ($g^{(2)} \to 1$).

\begin{figure}
	\includegraphics[width=.4\linewidth]{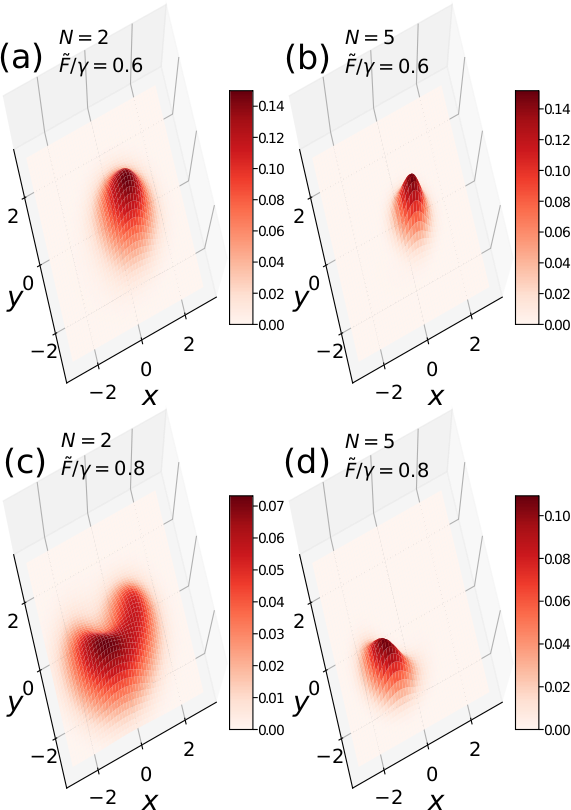} 
	\caption{\textbf{Rescaled Q-functions for local dissipation} below ($\tilde F/\gamma = 0.6$) and above ($\tilde F/\gamma = 0.8$) the bistability threshold for two values of $N$. The axes are $(x,y)=(\text{Re } \alpha, \text{Im }\alpha)/\sqrt{N}$. All the other parameters are the same as in Fig. \ref{fig: Bistability plots}.}
	\label{fig: Qfunc local dissipation bistability}
\end{figure}

An alternative way of ascertaining the presence of quantum bistability is through a quasiprobability distribution in the phase space. In Fig. \ref{fig: Qfunc local dissipation bistability} we plot the (rescaled) Q-function $Q(\tilde \alpha)= \bra{\tilde \alpha}\hat \rho_\text{ss} \ket{\tilde \alpha}/\pi$ under local dissipation for two values of $N$ and two pump values, one below [(a) and (b)] and one above ([(c) and (d)] the bistability threshold. Below threshold only one peak is observed, while above threshold a transition from a dark to a bright state begins at $N=2$, to reach completion for $N=5$.

\begin{figure}
  \includegraphics[width=1\linewidth]{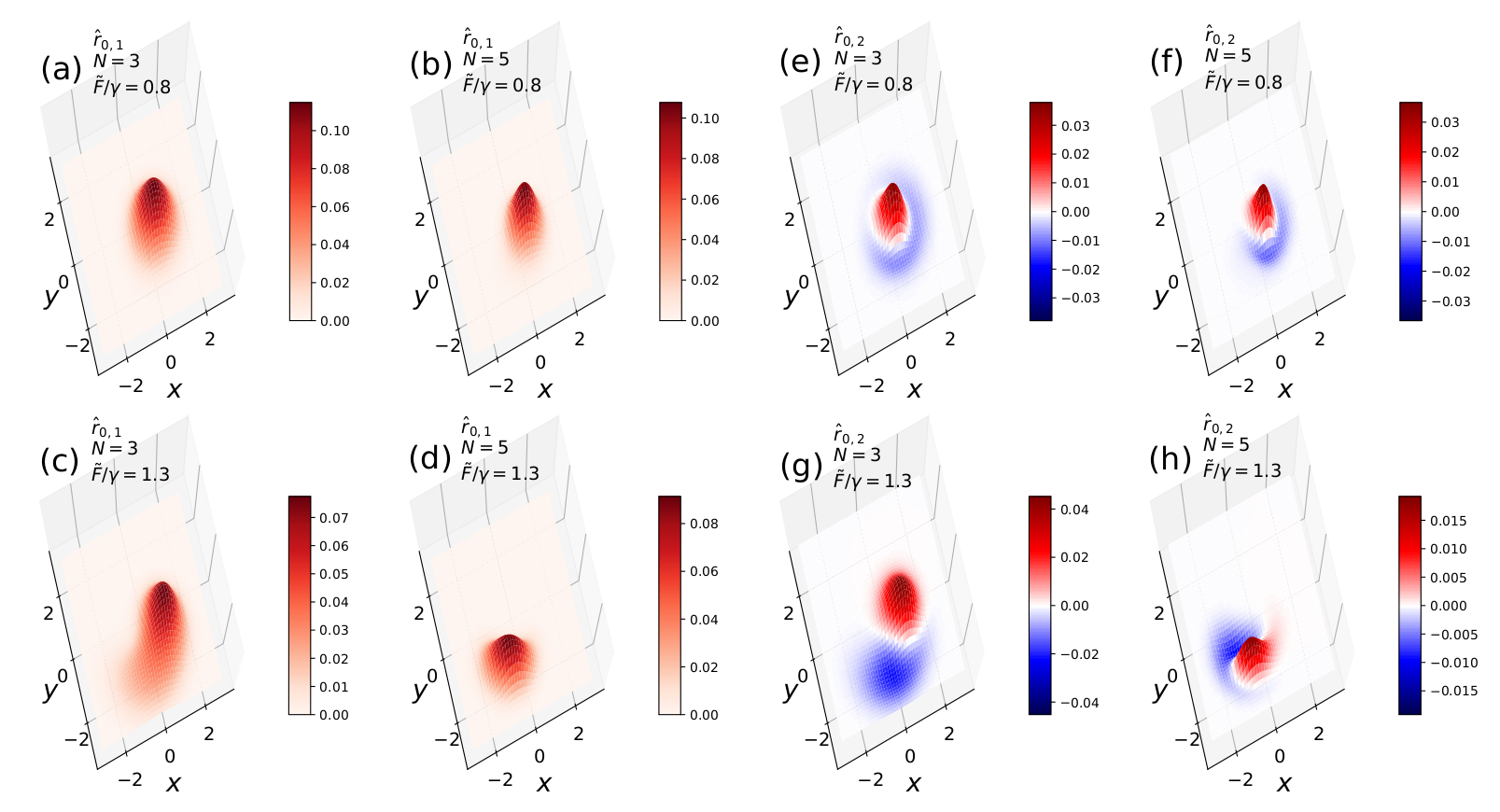}
\caption{\textbf{Rescaled Q-functions} corresponding to the two zero-eigenvalue eigenoperators of $\L$ that lead to the steady state \textbf{for nonlocal dissipation}. The steady state has the form of Eq. (\ref{equ: steady state with one conserved quantity}), with $c_{0,2} = \Tr[\hat Z_2 \hat \rho(0)]$. Here $\hat r_{0,1}$ (a-d) is a density matrix while $\hat r_{0,2}$ (e-h) is not. The Q-functions are plotted for two values of $N$, and two pump amplitudes: one below ($\tilde F/\gamma = 0.8$) and one above ($\tilde F/\gamma = 1.3)$ threshold. The axes are $(x,y)=(\text{Re } \alpha, \text{Im }\alpha)/\sqrt{N}$. All the other parameters are the same as in Fig. \ref{fig: non-local Liovillian gap} (and Fig. 1 in the main text).}
\label{fig: Qfunc bistability non-local}
\end{figure}

We now proceed to calculate the Q-function corresponding to the steady state reached when the dissipation is nonlocal. In this case, the steady state has the form of Eq. (\ref{equ: steady state with one conserved quantity}) with the conserved operator $\hat O = \hat Z_2$. Since the eigenvalues of $Z_2$ are $\pm 1$, the expectation value $\braket{\hat{Z}_2}$ for any density matrix is bounded between $-1$ and $+1$. In Fig. \ref{fig: Qfunc bistability non-local} we plot the rescaled Q-function corresponding to $\hat r_{0,1}$ [(a-d)] and $\hat r_{0,2}$ [(e-h)], for increasing values of $N$ and two pump strengths, one below ($\tilde F/\gamma = 0.8$) and one above ($\tilde F/\gamma = 1.3$) the bistability threshold. Any initial condition will lead to a linear combination of these two Q-functions in the steady state. The Q-function of $\hat r_{0,1}$ shows a very similar behavior to the one for local dissipation in Fig. \ref{fig: Qfunc local dissipation bistability}. The Q-function of $\hat r_{0,2}$, instead, takes also negative values in some regions (depicted in blue color) since it does not correspond to a density matrix.

As one can deduce from the plots for $N=3$ and $\tilde F/\gamma = 1.3$, when $\hat \rho_\text{ss} = \hat r_{0,1} + \hat r_{0,2}$, the Q-function of $\hat r_{0,2}$ will add some occupation probability to the dark state, while subtracting some from the bright state, in such a way that the transition will be observed for larger $N$. If instead one selects $\hat \rho_\text{ss} = \hat r_{0,1} - \hat r_{0,2}$, the transition will be accomplished for lower $N$. Interestingly, however, the participation of the Q-function corresponding to $\hat r_{0,2}$ fades away with increasing $N$ for high pump values, suggesting that in the thermodynamic limit the choice of initial condition for $c_{0,2}$ is unimportant above threshold. The initial condition will be important, however, for other eigenmodes of gaps that close in the thermodynamic limit, as well as for the eigenmodes associated to the limit cycles discussed in the main text.

\begin{figure}
 \centering
\includegraphics[width=0.6\textwidth]{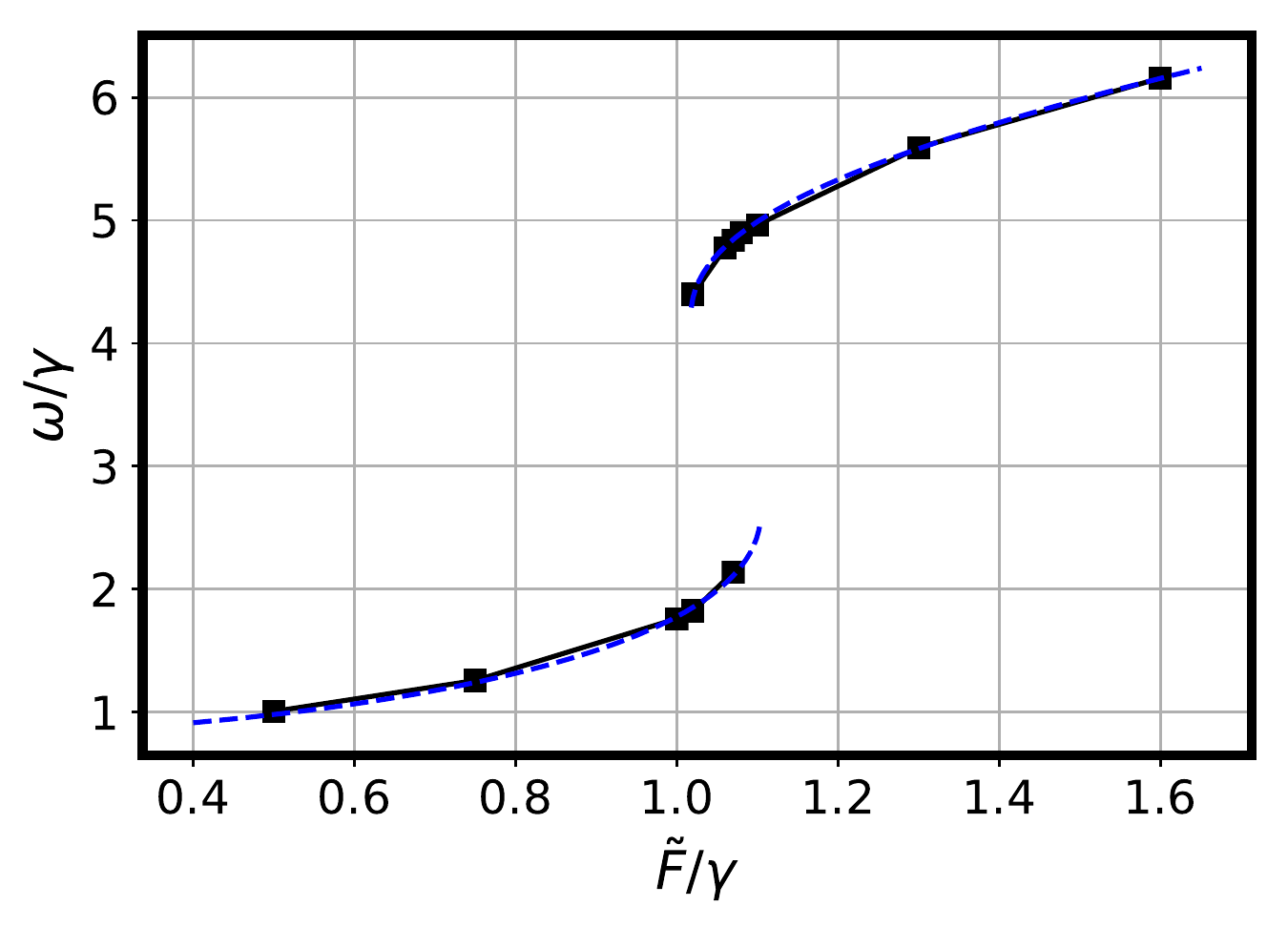}
\caption{\textbf{Semiclassical limit-cycle frequency as a function of the rescaled pump amplitude.} Black squares are calculated by time-integrating Eq. (\ref{equ: Gross-Pitaevskii equations rescaled}), also shown in Fig. 1(c) in the main text. The dashed blue line is the effective frequency given in Eq. (\ref{equ: effective LC frequency}) after the decoupling approximation. Same parameters as in Fig. \ref{fig: non-local Liovillian gap} (and Fig. 1 in the main text).}
\label{fig: effective vs full}
\end{figure}

\section{Effective antibonding decoupling}
The mechanism responsible for the emergence of limit cycles and the time-crystalline behaviour is an effective decoupling between the bonding and antibonding mode. As we have discussed in the main text, when the antibonding mode population is sufficiently small such that $|\tilde \alpha_A|^2 \equiv |\tilde \alpha_1-\tilde \alpha_2|^2/2 \ll 1$, bonding and antibonding modes are effectively decoupled in the semiclassical equations of motion. 

In order to appreciate this decoupling we start by writing the semicassical equations for the bonding and antibonding modes:
\begin{equation} \label{equ: bonding and antibonding SC equations}
\begin{split}
i\partial_t \tilde \alpha_B &= (-\Delta - J - i\gamma)\tilde \alpha_B + \frac{\tilde U}{2}(|\tilde \alpha_B|^2 \tilde \alpha_B + 2\tilde \alpha_A^2 \tilde \alpha_B^* + 4|\tilde \alpha_A|^2\tilde \alpha_B) + \sqrt{2}\tilde F \\
i\partial_t \tilde \alpha_A &= (-\Delta + J)\tilde \alpha_A + \frac{\tilde U}{2}(|\tilde \alpha_A|^2 \tilde \alpha_A + 2\tilde \alpha_B^2 \tilde \alpha_A^* + 4|\tilde \alpha_B|^2 \tilde \alpha_A).
\end{split}
\end{equation}
As the state of the system slowly approaches the stable steady states (shown in Fig. 1(a) in the main text), the antibonding mode $\tilde \alpha_A$ becomes small. Neglecting quadratic terms in $\tilde \alpha_A$ decouples the two equations \ref{equ: bonding and antibonding SC equations}. At the same time, the antibonding evolution becomes linear in $\tilde \alpha_A$ and the bonding mode enters as an energy-renormalization term. Solving the linear antibonding mode equation (and its complex conjugate) gives two eigenvalues that correspond to the oscillation frequency of the limit cycles:
\begin{equation} \label{equ: effective LC frequency}
\omega_\pm = \pm \sqrt{(-\Delta+J +2\tilde U|\tilde \alpha_B|^2)^2- \tilde U^2 |\tilde \alpha_B|^4},
\end{equation}
which are always real for the parameters we use in our work [$(J-\Delta)/\gamma = 0.8>0$]. In Fig. \ref{fig: effective vs full} we compare the limit-cycle frequencies (squares) shown in Fig. 1(c) in the main text, with the effective frequency (dashed line) obtained in Eq. (\ref{equ: effective LC frequency}) using the stable steady state solutions of Eq. (\ref{equ: Gross-Pitaevskii equations rescaled}) to evaluate $\tilde \alpha_B$. The agreement is perfect.

\section{Emergence of the bistable time crystal in single quantum jump trajectories}

\subsection{Local dissipation amplitude bistability}

\begin{figure}
  \includegraphics[width=0.4\linewidth]{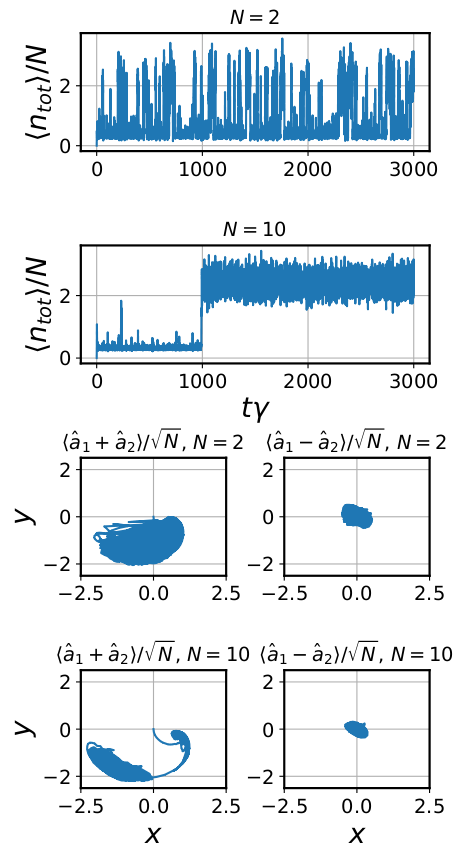} 
  \caption{\textbf{Single quantum trajectory for local dissipation at the bistability threshold} ($\tilde F/\gamma = 0.68$) for two values of $N$. We show the rescaled total number of bosons as a function of time at the top, and the rescaled bonding and antibonding modes evolution in phase space at the bottom. In the latter, the axis are $(x,y)= (\text{Re}(\alpha_1 \pm \alpha_2), \text{Im}(\alpha_1 \pm \alpha_2))\sqrt{N}$. All the other parameters are the same as in Fig. \ref{fig: Bistability plots}.}
\label{fig: single trajectory local dissipation}
\end{figure}

An important consequence of bistability is that in single experimental realisations, the system spends long times in either of two metastable states \cite{GarrahanMetastability}, switching between them with a rate much smaller than all other times scales in the system. This can be studied calculating single quantum jump Monte-Carlo trajectories \cite{PlenioRevMod} for pump values inside the bistability threshold. In \ref{fig: single trajectory local dissipation} we show a typical single trajectory for a BHD with local dissipation at the bistability threshold ($\tilde F/\gamma = 0.68$) for two values of $N$. In the top panel, we observe that as $N$ increases, the rescaled total number of bosons starts to fluctuate around two \textit{metastable} values, switching between them at a rate which is much smaller than the system decay rate (for $N=10$ the switching rate is $\sim 10^{-3}\gamma$). The metastable states are associated with the dark and bright states of amplitude bistability. This can be seen in the time evolution of the bonding mode in phase space shown at the bottom of the figure. The anti-bonding mode, instead, fluctuates around zero due to the swapping symmetry of the configuration, but the fluctuations decrease in size with growing $N$.


\begin{figure}
  \includegraphics[width=0.5\linewidth]{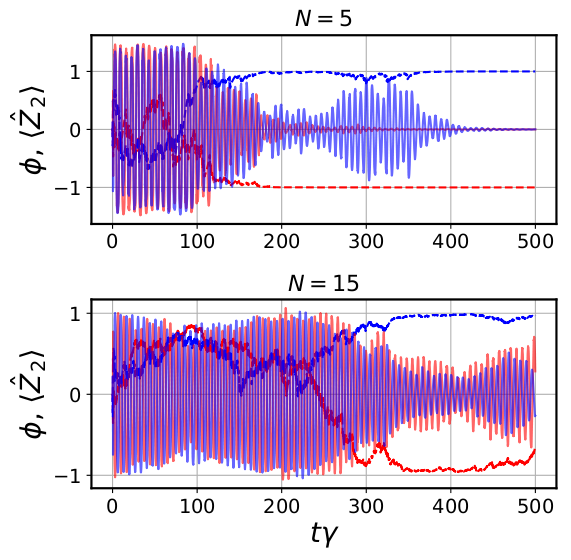} 
  \caption{\textbf{Dynamical evolution in single quantum trajectories for nonlocal dissipation.} Relative phase $\phi = \text{Arg}[\langle \hat a_1 \rangle] - \text{Arg}[\langle \hat a_2 \rangle]$ (solid line) and $\langle \hat Z_2 \rangle$ (dashed line) as a function of time for two values of $N$ and $\ket{\psi(0)} = \ket{0,1}$. Same parameters as in Fig. \ref{fig: non-local Liovillian gap} (and Fig. 1 in the main text).}
\label{fig: oscillations}
\end{figure}

\subsection{Nonlocal dissipation}

In the case of nonlocal dissipation, both the emergence of the time crystal and of bistability can be observed in single quantum trajectories.

Knowing that the swapping operator $\hat Z_2$ commutes with the BHD Hamiltonian and the nonlocal Lindblad operator $\hat a_1 + \hat a_2$, one can easily verify the following statement: if at any time $t=\tau$ during the quantum jump Monte-Carlo evolution the pure state $\ket{\psi(t)}$ becomes either symmetric or antisymmetric under the action of $\hat Z_2$, such that $\hat Z_2\ket{\psi(\tau)} = \pm \ket{\psi(\tau)}$, then {\bf(i)} for all consecutive times $t\ge \tau$ the expectation value of $\hat Z_2$ remains the same, and {\bf(ii)} the expectation value of the anti-bonding mode $(\hat a_1-\hat a_2)/\sqrt{2}$ is zero \textemdash{that} is,
\begin{equation}
 \bra{\psi(t)}\hat Z_2 \ket{\psi(t)} = \pm 1, \quad \bra{\psi(t)}(\hat a_1 - \hat a_2 )\ket{\psi(t)} = 0,
\end{equation}
for $t\ge \tau$. 

An initial state $\ket{\psi(0)}$ which is symmetric or anti-symmetric with respect to $\hat Z_2$, will remain so at all times. If, however, one selects some other initial state without the symmetry prescribed above, then the stochastic evolution will eventually evolve into the block Hilbert subspaces of symmetric or antisymmetric states. This is shown in Fig. \ref{fig: oscillations}, where we plot the relative phase $\phi$ and $\langle \hat Z_2 \rangle$ (dashed line) for two single quantum trajectories (drawn in different colors) which evolve into different subspaces. Here we have considered two values of $N$ and a low pump value ($\tilde F/\gamma = 0.4$). 

We can also note that the approach to the symmetric and antisymmetric block subspaces, for the pure state in a quantum trajectory, is delayed with increasing $N$. As long as this moment is yet to be attained, the trajectory evidences Josephson-like oscillations. This trend is the precursor of limit cycles as we get close to the $N \to \infty$ limit.

\begin{figure}
  \includegraphics[width=1.\linewidth]{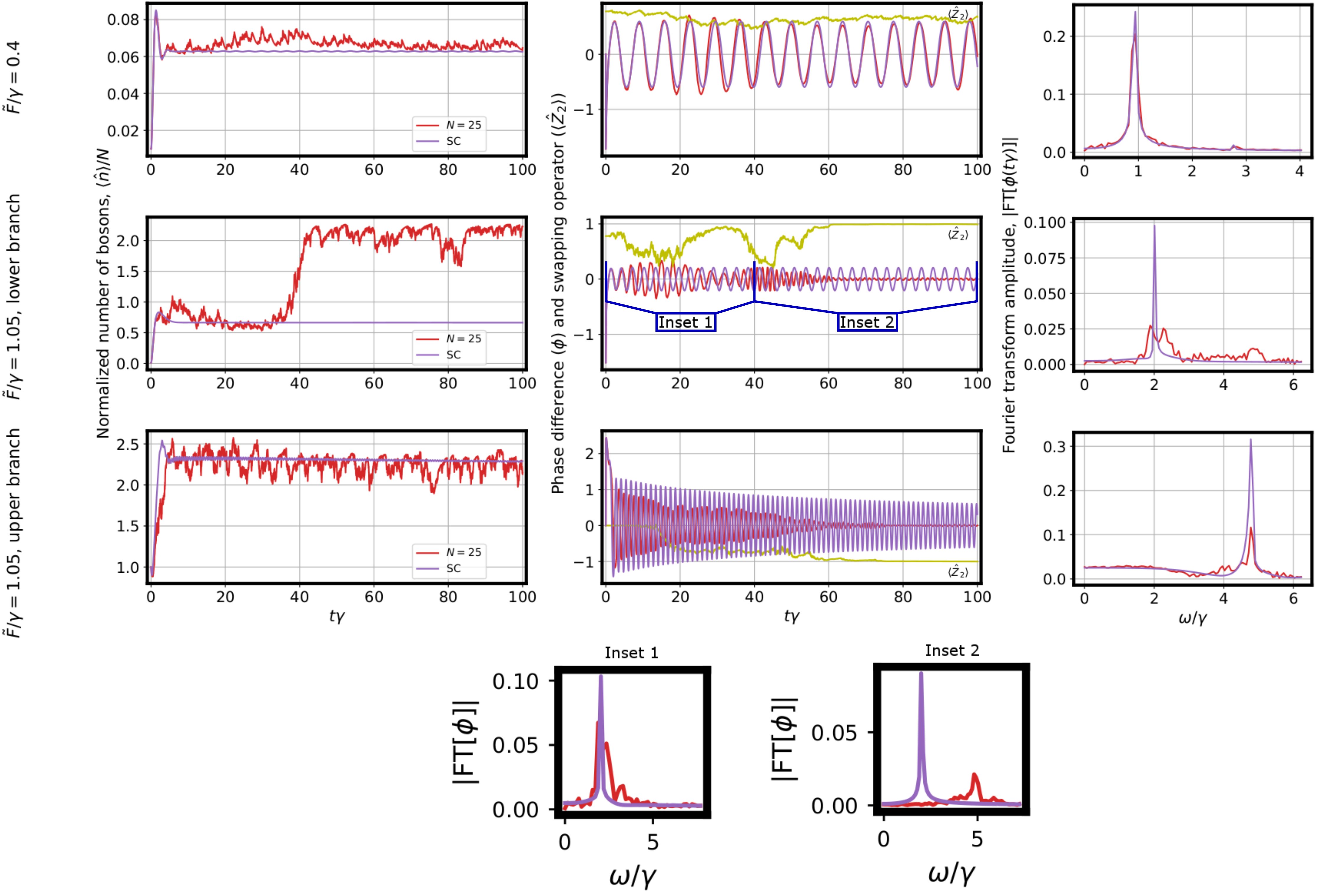} 
  \caption{\textbf{Single quantum trajectory and semiclassical trajectory} for two pump values far below (upper panel) and inside (middle and lower panels) the bistability region. The initial states in the upper and middle panels is $\ket{\psi(0)} = \ket{\alpha_1=0,\alpha_2=0.1\sqrt{N}}$, and in the lower panel is $\ket{\psi(0)} = \ket{\alpha_1=0,\alpha_2=\sqrt{N}}$. For $\tilde F/\gamma=1.05$, these initial states lead respectively to the lower and upper semiclassical branches (both in amplitude and frequency). All the other parameters are the same as in Fig. \ref{fig: non-local Liovillian gap} (and Fig. 1 in the main text).}
\label{fig: LC emergence trajectories}
\end{figure}

In Fig. \ref{fig: LC emergence trajectories} we show the expectation value of the normalized total number of bosons $\hat n/N$, the phase difference $\phi$, the expectation value of the swapping operator $\hat Z_2$, and the Fourier transform of $\phi$, for a single quantum trajectory and the semiclassical trajectory for different pump values and initial conditions. In the upper panel we consider a pump amplitude far below the bistability region ($\tilde F/\gamma = 0.4$) and a normalized initial condition $\ket{\psi(0)} = \ket{\alpha_1=0, \alpha_2=0.1 \sqrt{N}}$ (in the semiclassical trajectory this corresponds to $\tilde \alpha_1=0$ and $\tilde \alpha_2=0.1$). We can clearly observe that $\phi$ oscillates with a neat time-crystalline frequency even in the individual quantum trajectories.

In the middle and lower panels of Fig. \ref{fig: LC emergence trajectories} we show the same as in the upper panel but for an amplitude $\tilde F/\gamma = 1.05$ in the bistability region for two different initial conditions $\ket{\alpha_1=0, \alpha_2=0.1\sqrt{N}}$ and $\ket{\alpha_1=0, \alpha_2=\sqrt{N}}$, which lead (in the semiclassical evolution) to limit cycles in the lower and upper frequency branches, respectively. In the middle panel we have chosen a quantum trajectory that presents an amplitude switch (from the dark to the bright state) at $t\gamma \sim 40$ as can be seen in the plot for $\langle \hat n \rangle/N$. We show in the insets of the middle column that, when the amplitude switch occurs, the frequency of oscillations in $\phi$ switches between two values (from $\omega/\gamma \sim 2$ to $\omega/\gamma \sim 4.7$) which correspond to the lower and upper frequency branches of Fig. \ref{fig: effective vs full}. This means that the known bistability amplitude switching in single quantum trajectories is accompanied, in our model, by a frequency switching. We should mention that this is the general case of complex-amplitude bistability.

\begin{figure}
  \includegraphics[width=0.4\linewidth]{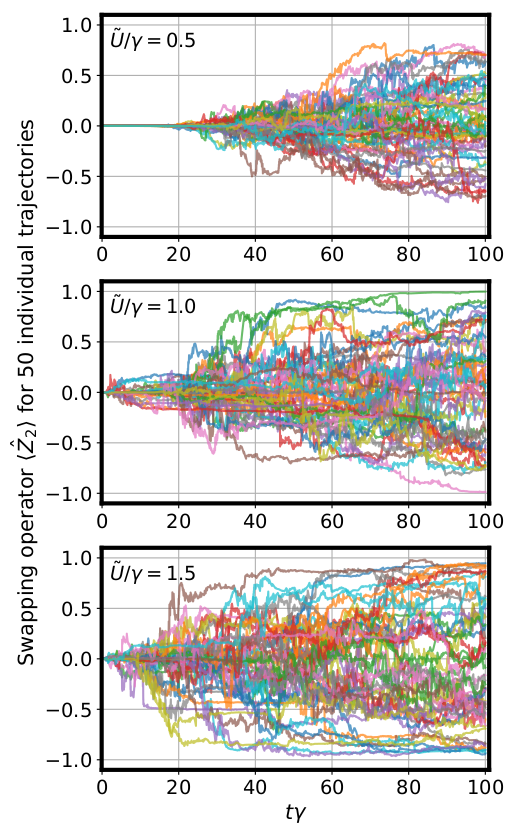} 
  \caption{\textbf{Swapping operator expectation value as a function of time for three different values of $\tilde U$} for 50 individual quantum jump trajectories. For larger $\tilde U$ the trajectories converge faster to the symmetry subspaces of $\langle \hat Z_2 \rangle = \pm 1$, giving shorter-lived oscillations. The initial state is $\ket{\psi(0)} = \ket{\alpha_1=0, \alpha_2=\sqrt{N}}$. The pump amplitude is $\tilde F/\gamma = 0.4$ and $N=25$. All the other parameters are the same as in Fig. \ref{fig: non-local Liovillian gap} (and Fig. 1 in the main text).}
\label{fig: different U}
\end{figure}

The effective decoupling between the bonding and antibonding modes takes place earlier or later in time depending on the value of the nonlinear coefficient $\tilde U$. A larger value means the two modes are more strongly coupled, thus oscillations decay faster for finite $N$. We show this in Fig. \ref{fig: different U} for $\tilde U\gamma = 0.5$, $1.0$ and $1.5$. We can see that for smaller $\tilde U$ the single trajectories take longer times to reach the symmetry subspaces of $\hat Z_2$, which means that the Josephson-like oscillations live longer. If $\tilde U \to 0$, oscillations last forever as the antibonding mode is decoupled from the bonding mode and evolves coherently, as we have explained in the main text as well as here.

\section{Microscopic derivation of the Lindblad Master Equations}

We consider the system Hamiltonian
\begin{equation} \label{equ: BH Hamiltonian lab frame}
\begin{split}
 \hat H = \sum_i &\omega_c \hat a^\dag_i \hat a_i + U \hat a^\dag_i \hat a^\dag_i \hat a_i \hat a_i -J(\hat a^\dag_{i} \hat a_{i+1} + \hat a_{i} \hat a_{i+1}^\dag) \\
&+ F_{i}(e^{-i\omega_pt} \hat a^\dag_i + e^{i\omega_pt} \hat a_i),
\end{split}
\end{equation}
in the laboratory frame and a bosonic environment $\hat H_B = \sum_{k} \omega_k  \hat b_k^\dag \hat b_k$ initially at thermal equilibrium, described by the density matrix $\hat \rho_B = e^{-\beta \hat H_B}/Z_B$. We assume a linear system-environment coupling, that, in its more general form, can be described by the Hamiltonian $\hat H_I = \sum_{k,i} (\gamma_{k,i} \hat a_i \hat b_k^\dag +\text{h.c.})$. Note, however, that by virtue of symmetry, the coefficients $\gamma_{k,i}$ should be equal for both sites, i.e., $\gamma_{k,1} = \gamma_{k,2} \equiv \gamma_{k}$. We move to a frame rotating with frequency $\omega_p$ using the unitary operator
\begin{equation}
 \hat U_r(t) = \exp\left[it\omega_p \left(\sum\limits_{i=1,2} \hat a_i^\dag \hat a_i + \sum_k \hat b_k^\dag \hat b_k\right) \right].
\end{equation}
This operation transforms the total Hamiltonian $\hat H_\text{tot} = \hat H + \hat H_B + \hat H_I$ into $\hat{\mathcal{H}}_\text{tot} = \hat{\mathcal H} + \hat{\mathcal H}_B + \hat{\mathcal H}_I$, where
\begin{equation}
\begin{split}
 \hat{\mathcal H}_B =& \sum_{k} (\omega_k-\omega_p)\hat b_k^\dag \hat b_k, \\ \hat{\mathcal H}_I =& \hat H_I = \sum_k \gamma_k (\hat a_1 + \hat a_2) \hat b_k^\dag + \text{h.c.},
\end{split}
\end{equation}
and $\hat{\mathcal H}$ is given in Eq. (2) in the main text.

As a second step we perfom the Born-Markov approximations in the interaction picture. Defining the operator $\hat a \equiv \hat a_1 + \hat a_2$ for simplicity, one obtains \cite{BreuerBook}
\begin{equation}
 \begin{split}
  \partial_t \tilde \rho(t) &= - \int\limits_0^\infty ds \Tr_B[\tilde{\mathcal H}_I(t),[\tilde{\mathcal H}_I(t-s), \tilde \rho(t)  \otimes \hat \rho_B] ] \\
&= \int\limits_0^\infty ds \biggl\{ C_1(s) [\tilde a^\dag(t-s) \tilde \rho(t) \tilde a(t) - \tilde a(t) \tilde a^\dag(t-s) \tilde \rho(t)] \\
& \quad \qquad + C_2(s) [\tilde a(t) \tilde \rho(t) \tilde a^\dag(t-s) - \tilde \rho(t) \tilde a^\dag(t-s) \tilde a(t)] \biggr\} \\
&\,\,\, + \text{h.c.},
 \end{split}
\end{equation}
where $\tilde O(t) = e^{it(\hat{\mathcal H} + \hat{\mathcal H}_B)} \hat O(t) e^{-it(\hat{\mathcal H} + \hat{\mathcal H}_B)}$ is the time evolution of the operator $\hat O(t)$ in the interaction picture. The environment correlation functions read
\begin{equation}
\begin{split}
 C_1(s) &= \sum_k |\gamma_k|^2 \Tr_B[ \tilde b_k^\dag (s) \tilde b_k(0) \hat \rho_B] \\
&= \sum_k |\gamma_k|^2 e^{is(\omega_k-\omega_p)} n_B(\omega_k), \\
 C_2(s) &= \sum_k |\gamma_k|^2 \Tr_B[ \tilde b_k (0) \tilde b_k^\dag(s) \hat \rho_B] \\
&= \sum_k |\gamma_k|^2 e^{is(\omega_k-\omega_p)} [1 + n_B(\omega_k)],
\end{split}
\end{equation}
with $n_B(x)=(e^{\beta x} -1)^{-1}$ the Bose-Einstein distribution.

We now make a third approximation that is an alternative to the standard rotating wave (secular) approximation \cite{Marti2017}. The environment correlation functions decay on a small time scale $\tau_B$, which is the assumption underlying the Markovian approximation. We require that for shorter times ($s\leq \tau_B$) the condition $\tilde a^\dag(t-s) \backsimeq \tilde a^\dag(t)$ is satisfied, so that we can integrate only the environment correlation functions. This is only possible if the environment correlation time is much smaller than the smallest time scale of our system, which is, roughly, $\tau_B \times \max\{|\Delta|, U, -J, F_{i}\} \ll 1$. With this condition in hand, and assuming that the environment has a vanishing thermal occupation, we obtain in the Schr\"{o}dinger picture the Lindblad master equation for nonlocal dissipation Eq. (1) in the main text, with a decay rate $\gamma = 2\pi \mathcal{S}(\omega_p)$ where 
\begin{equation}
\mathcal S(\omega)=\sum_k |\gamma_k|^2 \delta(\omega-\omega_k) 
\end{equation}
is the environment spectral density. We have ignored the Lamb-Shift $\hat{\mathcal H}_{LS}=\Omega(\omega_p) (\hat a_1^\dag \hat a_1 + \hat a_2^\dag \hat a_2)$, with $\Omega(\omega_p) = \mathcal{P} \int_0^\infty d\omega \mathcal{S}(\omega) /(\omega_p-\omega)$, as it can be absorbed by the on-site detuning energy $\Delta$ of the Hamiltonian. Here, $\mathcal{P}$ stands for the \textit{principal value}.

The third approximation we have made above can be expressed as $C_2(s) \backsimeq C_2(s) e^{\pm i\sigma s}$ $\forall s$, with $\sigma \equiv \max\{|\Delta|, U, -J, F_{i}\}$. Noting that $C_2(s) = \int_{-\infty}^{\infty} d\omega \, \mathcal S(\omega)e^{i(\omega-\omega_p)}$, this is fulfilled as long as the spectral density does not vary appreciably around $\omega_p$ in an interval of width $2\sigma$ \cite{Marti2017}: 
\begin{equation}
 |\mathcal S(\omega_p \pm \sigma) - \mathcal S(\omega_p)| \ll \mathcal S(\omega_p).
\end{equation}
This condition is satisfied for Ohmic spectral densities as long as $\omega_p \gg \sigma$.

Strictly speaking, because our system has an unbounded Hamiltonian, the condition $\tau \sigma \ll 1$ may not be enough to safely carry out this approximation. To be more careful, let us proceed to the following definition:
\begin{equation}
 \mathbb{\Omega} := \{ \omega= \epsilon_j-\epsilon_i | \bra{\epsilon_i}\hat \rho(t)\ket{\epsilon_j} \approx 0 \, \forall t \}, 
\end{equation}
where $\{\ket{\epsilon_i}, \epsilon_i\}$ are the eigenkets and eigenenergies of the Hamiltonian $\hat{\mathcal H}$. If $\bar{\omega} = \max\{ |\bar{\mathbb{\Omega}}| \}$, where $\bar{\mathbb{\Omega}} = \{\text{all frequencies in the system}\}/\mathbb{\Omega}$, then, our condition should read
\begin{equation}
 s\bar{\omega} \ll 1 \quad \text{for } s\lesssim \tau_B.
\end{equation}
This maximum frequency $\bar \omega$ is the highest transition frequency occurring in the dynamic evolution of our system, or, equivalently, the inverse of the smallest time scale characterizing the system response.

Two comments are in order here. The first is that for the derivation of LME with local dissipation, the same procedure outlined here can be used, but one must consider two identical environments, say $\mathcal B_1$ and $\mathcal B_2$, independently coupled to system modes 1 and 2, respectively. The Hamiltonian operators describing the environments would now read $\hat H_{B_1} = \sum \omega_k \hat b_{k,1}^\dag \hat b_{k,1}$ and $\hat H_{B_2} = \sum \omega_k \hat b_{k,2}^\dag \hat b_{k,2}$, where $\hat b_{k,1}$ and $\hat b_{k,2}$ are bosonic operators acting on different environment Fock spaces, i.e., $[\hat b_{j,1}, \hat b_{k,2}] = [\hat b_{j,1}, \hat b_{k,2}^\dag]=0$, while the system-environment couplings should read $\hat H_{I_1} = \sum_k (\gamma_k \hat a_1 \hat b_{k,1}^\dag + \text{h.c.})$ and $\hat H_{I_2} = \sum_k (\gamma_k \hat a_2 \hat b_{k,2}^\dag + \text{h.c.})$, respectively. This is, to the best of our knowledge, the only possible way to derive LME for local dissipation with bosonic baths and linear system-bath couplings. The alternative input-output formalism, or the equivalent collisional models \cite{Ciccarello2017}, make use of two identical reservoirs alongside the stringent constraint of a constant spectral density.

The second comment concerns the third approximation we used to derive the Lindblad equation. This approximation breaks down in the thermodynamic limit we have considered, for which the energy scales with a parameter $N$ approaching infinity. This implies that there are time scales in our system smaller than the decay times of bath correlations. We can still, however, probe the system response when one approaches that limit while still remaining under the regime where the approximation is valid. Outside this regime, instead of the third approximation outlined above, one should use the rotating wave (secular) approximation, where fast oscillatory terms are neglected and the system relaxes to thermal equilibrium.

\subsection{Imperfections in the symmetric coupling}

As we concluded in the main text, the primary difficulty in implementing the configuration of nonlocal dissipation lies on the fact that the two modes of the system may couple with the same strength to the single environment, yet the complex phases of the coupling constants, $\gamma_{k,i}$, may differ between $i=1, 2$. In the following we briefly discuss this possibility.

Let us write the coupling constants in the form $\gamma_{k,i} = \gamma_{k} e^{i\phi_{k,i}}$, defining the difference between the two phases as $\Delta \phi_k \equiv \phi_{k,1} - \phi_{k,2}$. In this case, the LME in the Schr\"{o}dinger picture reads
\begin{equation}
\partial_t \hat \rho = -i\left[\hat{\mathcal H} + \sum\limits_{ij} \Omega_{ij}(\omega_p) \hat a_j^\dag \hat a_i, \hat \rho \right] + 2\pi \sum_{ij} S_{ij}(\omega_p) \left( \hat a_i \hat \rho \hat a_j^\dag - \frac{1}{2} \{\hat a_j^\dag \hat a_i, \hat \rho \} \right),
\end{equation}
with $\Omega_{ij}(\omega_p) = -\mathcal P \int\limits_0^\infty d\omega \, \frac{S_{ij}(\omega)}{\omega - \omega_p}$, and $S_{ij}(\omega) = \sum_k |\gamma_k|^2 e^{i(\phi_{k,i} - \phi_{k,j})} \delta(\omega_k-\omega)$. We note that the first-order (in $\Delta\phi_k$) contribution to the Lamb-Shift vanishes; therefore, only the zero-order term remains which can be absorbed into the detuning $\Delta$, which is what we have done before. In order to recast the dissipation in diagonal form, we can define a new operator
\begin{equation} \label{equ: new Lindblad operator}
\hat c_k = \frac{1}{\sqrt{2}} (e^{i\Delta \phi_k/2} \hat a_1 + e^{-i \Delta \phi_k/2} \hat a_2) = \cos(\Delta \phi_k/2) \hat a_B  + i \sin(\Delta \phi_k/2) \hat a_A,
\end{equation}
obtaining the corresponding dissipator
\begin{equation} \label{equ: new Lindblad equation}
\mathcal D(\hat \rho) = 4\pi \sum_k |\gamma_k|^2 \delta(\omega_k - \omega_p) \left( \hat c_k \hat \rho \hat c_k^\dag - \frac{1}{2}\{ \hat c_k^\dag \hat c_k,\hat \rho\} \right).
\end{equation}

\begin{figure}
	\includegraphics[width=1\linewidth]{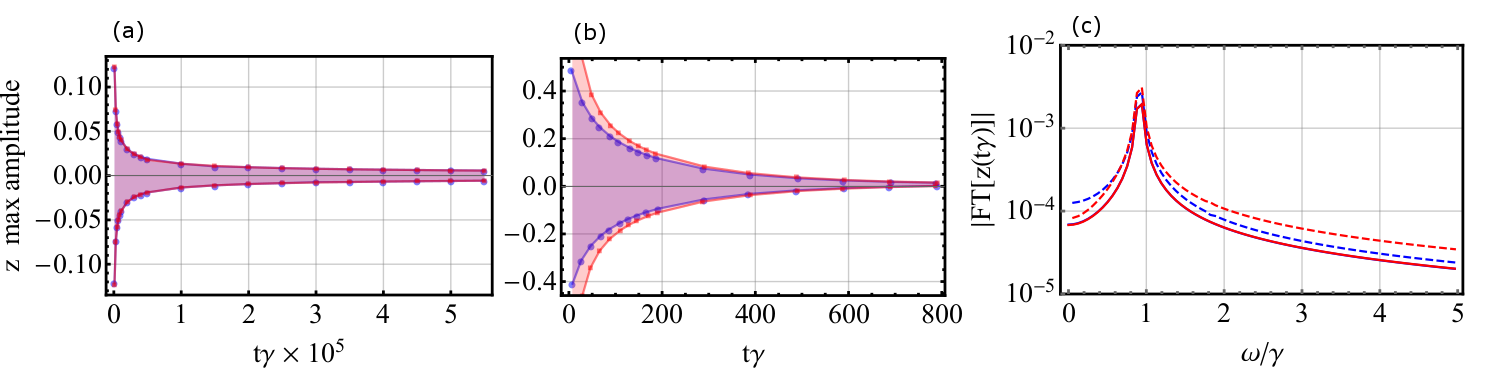} 
	\caption{\textbf{Long-lived oscillations for perfect and imperfect nonlocal dissipation.} Frames {\bf(a)} and {\bf(b)} show the envelope of the oscillating population difference $z = |\tilde{\alpha}_1|^2-|\tilde \alpha_2|^2$ for perfect and imperfect nonlocal dissipation, respectively. Two random initial conditions were chosen, and are depicted in different colors. Frame {\bf(c)} shows a Fourier transformation of $z$ in a time window of $100\gamma^{-1}$ at the end of the evolution shown in (a) and (b). Continuous (dashed) lines correspond to perfect (imperfect) dissipation. The pump is $\tilde F/\gamma = 0.4$, and all the other parameters are same as in Fig. 1 in the main text.}
	\label{fig: robustness limit cycles}
\end{figure}

The new Lindblad operator in Eq. (\ref{equ: new Lindblad equation}) does not commute with $\hat Z_2$, hence it breaks explicitly the continuous swapping symmetry. Due to the presence of the Dirac delta-function, however, the only relevant $k-$modes are those near the pump frequency $\omega_p$. This means that in order for the dissipation to break the symmetry there must be a nonnegligible phase difference $\Delta\phi_k$ for $\omega_k \approx \omega_p$. If that difference is small yet finite, the main contribution to the dissipation will be due to the bonding mode $\hat a_B$, as $\hat c_k \approx \hat a_B + i(\Delta\phi_k/2)\hat a_A$. Such a dissipator lifts the zero-eigenvalue degeneracy of the Liouvillian, and, as we have numerically verified, destroys also the semiclassical limit cycles. Long-lived oscillations can still be observed if $\Delta\phi_k$ is small enough. To illustrate this point, we study the long-time dynamics of the semiclassical equations of motion obtained from the new evolution $\partial_t \hat \rho = -i[\hat{\mathcal H}, \hat \rho] + 2\gamma \mathcal D[\hat c_p](\hat \rho)$, where $\gamma = 2\pi \sum_k |\gamma_k|^2 \delta(\omega_k-\omega_p)$ and $\hat c_p$ is given by Eq. (\ref{equ: new Lindblad operator}) for the mode $\omega_p$. The modified (rescaled) Gross-Pitaevskii equations read
\begin{equation} \label{ecc: modified GP equations}
\begin{split}
i \partial_t \tilde \alpha_1 &= (-\Delta - i\gamma/2 + 2 \tilde U|\tilde\alpha_1|^2) \tilde \alpha_1 -(J + i e^{-i\Delta \phi_p} \gamma/2) \tilde \alpha_2 + \tilde F, \\
i \partial_t  \tilde \alpha_2 &= (-\Delta -i\gamma/2 + 2 \tilde U| \tilde \alpha_2|^2) \tilde \alpha_2 - (J + i e^{i\Delta \phi_p} \gamma/2) \tilde \alpha_1 + \tilde F.
\end{split}
\end{equation}
The new terms $e^{\pm i \Delta \phi_p}$ introduced have real and imaginary parts. The imaginary part will by added to the coherent part of the dynamics and is relatively innocuous. What is really important is the real part, which adds a local dissipation contribution to the system. We choose $\Delta \phi_p = \cos^{-1}(0.99)$ to yield 1$\%$ ($99\%$) of local (nonlocal) dissipation, and compare in the long time-behaviour, as predicted in Eqs. (\ref{equ: Gross-Pitaevskii equations rescaled}) and (\ref{ecc: modified GP equations}) and depicted in Fig. \ref{fig: robustness limit cycles}. In Fig. \ref{fig: robustness limit cycles}(a) and (b) we show the envelope of the oscillating population difference $z(\tau) = |\tilde \alpha_1(\tau)|^2 - |\tilde \alpha_2(\tau)|^2$ for perfect and imperfect nonlocal dissipation, respectively. In each case we have chosen two random initial conditions, shown in different colors. For perfect nonlocal dissipation, $z$ slowly converge to its final oscillatory form at very large times ($t\gamma \sim 3.5 \times 10^5$), while for imperfect dissipation, $z$ goes to zero and the relaxation is much faster ($t\gamma \sim 800$); it is, however, very slow compared to the time scale set by decay rate $\gamma$, which means that the oscillatory behaviour could be observed in the laboratory even in the presence of imperfect nonlocal dissipation.

Another important aspect characterizing the inclusion of the perturbative terms is the period robustness. In Fig. \ref{fig: robustness limit cycles}(c) we show a Fourier transformation of $z$ in a time window of $100\gamma^{-1}$ at the end of the evolution shown in Fig. \ref{fig: robustness limit cycles}(a) and (b). In blue and red lines (dashed lines) we show the Fourier transformation for perfect (imperfect) nonlocal dissipation.  We can see that the frequency peak ($\omega/\gamma \approx 0.9424$) is the same in both cases, independently of the initial condition. A similar frequency robustness over driving imperfections has been experimentally observed in a \textit{Discrete Time Crystal} \cite{Choi2017}.

\bibliography{PRBSubmission}